\documentclass[pra,showpacs,floatfix,twocolumn,letterpaper,footinbib]{revtex4}

\usepackage{graphicx}
\usepackage{bm,amsmath}
\usepackage{dcolumn}
\usepackage{natbib}

\newcommand{\kms}{km~s$^{-1}\,$}
\newcommand{\ms}{m~s$^{-1}\,$}
\newcommand{\cms}{cm~s$^{-1}\,$}

\newcommand{\cmm}{cm$^{-3}\,$}
\newcommand{\daa}{$\Delta\alpha/\alpha\,$}
\newcommand{\dmm}{$\Delta\mu/\mu\,$}

\def\la{\;
\raise0.3ex\hbox{$<$\kern-0.75em\raise-1.1ex\hbox{$\sim$}}\; }
\def\ga{\;
\raise0.3ex\hbox{$>$\kern-0.75em\raise-1.1ex\hbox{$\sim$}}\; }

\begin{document}
\title{On spatial variations of the
electron-to-proton mass ratio in the Milky Way
}
\author{S.~A.~Levshakov$^{1}$}
\author{P.~Molaro$^{2}$}
\author{M.~G.~Kozlov$^{3}$}
\affiliation{$^1$Ioffe Physico-Technical Institute, Politekhnicheskaya Str. 26,
              194021 St. Petersburg, Russia}
\affiliation{$^2$INAF-Osservatorio Astronomico di Trieste, 
Via G. B. Tiepolo 11, 34131 I, Trieste, Italy}
\affiliation{$^3$Petersburg Nuclear Physics Institute, Gatchina, 188300, Russia}
\date{ \today }

\begin{abstract}
We aim at applying the ammonia method, recently proposed to explore the
electron-to-proton mass ratio $\mu = m_{\rm e}/m_{\rm p}$, to Galactic sources.
Emission lines due to ammonia are originated in the numerous 
interstellar molecular
clouds allowing to probe $\mu$ at different galactocentric distances.
High quality radio-astronomical observations of molecular cores
in lines of NH$_3$ $(J,K) = (1,1)$, CCS $J_N = 2_1-1_0$, HC$_3$N $J = 5-4$,
and N$_2$H$^+$ $J = 1-0$ are used to measure the relative radial
velocity offsets, $\Delta V$, between the NH$_3$ (1,1) 
and other molecular transitions.
Robust statistical analysis is applied to the $n = 207$ individual measurements 
in the Perseus molecular cloud (PC), the Pipe Nebula (PN), and the infrared
dark clouds (IRDCs) to deduce the mean value of $\Delta V$ and its uncertainty.  
The measured values of $\Delta V$ from the PC, PN, and
IRDCs\ show statistically significant positive velocity shifts between
the line centers of NH$_3$ and other molecules.
The most accurate estimates are obtained from carefully
selected subsamples of the NH$_3$/CCS pairs observed in the  
PC ($n = 21$) and the PN ($n = 8$),
$\Delta V = 36\pm7_{\rm stat}\pm13.5_{\rm sys}$ \ms\ and 
$53\pm11_{\rm stat}\pm13.5_{\rm sys}$ \ms, respectively, 
and from the $n = 36$ NH$_3$/N$_2$H$^+$ and 
$n = 27$ NH$_3$/HC$_3$N pairs observed in the IRDCs,
$\Delta V = 148\pm32_{\rm stat}\pm13.6_{\rm sys}$ \ms\ and 
$115\pm37_{\rm stat}\pm31_{\rm sys}$ \ms, respectively.  
Being interpreted in terms of the electron-to-proton mass ratio
variation, this gives \dmm $\sim (4-14)\times10^{-8}$,  
which is an order of magnitude more sensitive than previous
astronomical constraints on this quantity. 
If \dmm\ follows the gradient of the local
gravitational potential, as suggested in some scalar field models,
then the obtained result is in sharp 
contradiction ($\sim 5$ orders of magnitude)
with laboratory atomic clock experiments
carried out at different points in the gravitational potential of the Sun.
However, the measured signal might be consistent with chameleon-type scalar
field models which predict a strong dependence of masses and coupling
constants on the ambient matter density.  
The confirmation of a spatial variation of \dmm\ will 
help to distinguish between many
theoretical proposals suggested to explain the nature of dark energy. 
New control measurements of the velocity offsets
involving other molecules and a wider range of objects 
along with verification of rest frequencies 
are needed to investigate different instrument systematics and to
come to a more certain conclusion.  
\end{abstract}
\maketitle

\section{Introduction}
\label{Sect1}

A fundamental characteristic of the universe is that 
decelerating expansion changed to accelerating at redshift $z \sim 1$ 
\cite{RFC98,RKS99,PAG99}.
This late time acceleration is usually attributed to 
some negative-pressure component of the bulk
of energy density, referred to as `dark energy'.
The nature of dark energy remains unknown in spite of
a good deal of theoretical and experimental efforts \cite{CST06}.
The most popular scenario is
a dynamically evolving scalar field $\varphi$, named `quintessence' \cite{CDS98},
with the energy density subdominant at a matter dominated (decelerating)
epoch ($1 \la z \la 1000$), and dominant at lower redshifts \cite{PR03}.
Scalar field models allow to alleviate
the so-called `cosmic coincidence' problem~-- the fact that matter and
vacuum energy contribute by comparable amounts to the energy density
at the present cosmological epoch \cite{BBM01}.
The scalar field(s) coupling to ordinary matter leads 
unavoidably to long-range forces and variability of the physical
constants violating the Equivalence Principle. 
Thus, any laboratory or space-based 
experiments as well as astronomical studies aimed at 
probing the variations of the physical constants, are
the most important tools to test new theories against observations. 

Present experimental facilities allow us to probe the variability of the
fine-structure constant 
$\alpha = e^2/(\hbar c)$
and the electron-to-proton mass ratio
$\mu = m_{\rm e}/m_{\rm p}$, or different
combinations of the proton gyromagnetic ratio $g_{\rm p}$
with $\alpha$ and $\mu$  \cite{FK07}.

The most accurate laboratory constraint on the temporal variation of $\alpha$ of
$\dot{\alpha}/\alpha = (-1.6\pm2.3)\times10^{-17}$ yr$^{-1}$ 
is obtained from atomic clocks experiments \cite{RHS08}.
Being linearly extrapolated to redshift $z \sim 2$, or
$t \sim 10^{10}$ yr, this laboratory bound leads to 
\daa~= $(-1.6\pm2.3)\times10^{-7}$ which is well below 
a conservative upper limit on 
$|\Delta\alpha/\alpha| < 10^{-5}$ 
stemmed from the analysis of quasar absorption-line systems
\cite{MWF08,SCP08,MLM08}.
Here $\Delta \alpha/\alpha = (\alpha' - \alpha)/\alpha$, with $\alpha, \alpha'$
denoting the values of the fine-structure constant in the laboratory and the
specific absorption/emission line system of a galactic or extragalactic object
(the same definition is applied to \dmm).  
However, laboratory experiments and quasar absorption spectra investigate 
very different time scales and different regions of the universe, 
and the connection between them is somewhat model dependent 
\cite{MB04a,MB04b}.

The direct and model-free estimate of time variation of $\mu$ was deduced from 
comparison of the frequency of a rovibrational transition in SF$_6$ molecule
with the fundamental hyperfine transition in Cs:
$\dot{\mu}/\mu = (3.8\pm5.6)\times10^{-14}$ yr$^{-1}$ \cite{SBC08}.
Astrophysical estimates of $\mu$ at high redshifs 
are controversial in the claims of nonzero and zero changes of $\mu$:
$\Delta \mu/\mu = (-2.4\pm0.6)\times10^{-5}$ was inferred from
the analysis of the H$_2$-bearing clouds at 
$z = 2.6$ and $3.0$ \cite{RBH06}, 
but this value was not confirmed later \cite{WR08,KWM08}.
A better accuracy bound was obtained at lower
redshift $z = 0.68$ from the analysis of radio-frequency transitions in
NH$_3$, CO, HCO$^+$, and  HCN:
$\Delta \mu/\mu = (-0.6\pm1.9)\times10^{-6}$ \cite{FK07b},
which favors a non-variability of $\mu$ at the level of $\sim3\times10^{-6}$.

In these studies it was assumed that the rate of time variations 
dominates over the rate of possible spatial variations.
However, this may not be the case if scalar fields trace 
the gravitational field inhomogeneities \cite{B05,BM05}. 
The spatial variation of constants in a 
changing gravitational potential of the Sun were
studied in laboratory measurements of the atomic clock transition
$^1$S$_0$--$^3$P$_0$ in neutral $^{87}$Sr relative to the Cs standard and of
the radio-frequency $E1$ transition between nearly
degenerate, opposite-parity levels of atomic dysprosium (Dy). 
The values for the couplings of $\alpha$ and $\mu$
to the gravitational field were reported to be 
$k_\alpha = (2.3\pm3.1)\times10^{-6}$, 
$k_\mu = (-1.1\pm1.7)\times10^{-5}$ \cite{BLC08}, and 
$k_\alpha = (-8.7\pm6.6)\times10^{-6}$ \cite{FCL07}, respectively.

Thus, the current laboratory and astrophysical estimates 
of the temporal and/or spatial changes of the coupling constants and 
fermion masses 
do not exhibit any meaningful features which could be interpreted in terms
of scalar field effects. On the other hand, we do observe the late
time acceleration of the universe driven most probably by ultra-light 
scalar fields. 
To reconcile these results chameleon-type models were suggested  
\cite{KW04a,KW04b,GK04,BvB04,FN06,UGK06,OP08}
which allow scalar
fields to evolve on cosmological time scales today and to have 
simultaneous strong couplings of order unity to matter, as
expected from string theory. 
The key assumption in this scenario is that the mass of the scalar field 
depends on the local matter density.
This explains why cosmological scalar fields, such as quintessence,
are not detectable in local tests of the Equivalence Principle 
since we live in a dense
environment where the mass of the field can be sufficiently large and
$\varphi$-mediated interactions are short-ranged. On the other hand,
the cosmological matter density is about $10^{30}$ times smaller and thus
the mass of the scalar field can be very low, of order of $H_0$, the
present Hubble constant. This allows the field to evolve cosmologically
today.
Laboratory experiments aimed at detections of chameleon particles
are nowadays carried out at Fermilab \cite{CWB08}.

In this paper we explore the astrophysical implications of the scalar field
effect on the electron-to-proton mass ratio, $\mu$,
following suggestions on experimental tests of \dmm\
in our Galaxy as formulated in \cite{OP08}.
We consider the constraints on the spatial
variations \dmm\ caused by the relative Doppler shifts of molecular
transitions in cold molecular cores observed in the Milky Way disk at
different galactocentric distances.

Our assumptions and basic physical properties of dense molecular cores
which will be needed in the sequel are discussed in Sect.~\ref{Sect2}.
The statistical analysis of the line position
measurements in the Perseus molecular complex,
the Pipe Nebula, and the infrared dark clouds are presented in Sect.~\ref{Sect3}.
The obtained results are discussed in Sect.~\ref{Sect4}, and
conclusions are given in Sect.~\ref{Sect5}.

%-------------------------- Table 1
\begin{table}[t!]
\caption{Molecular transitions and $V_{\rm LSR}$ uncertainties,
$\varepsilon_v$, used in calculations of \dmm.
The numbers in parentheses correspond to $1\sigma$ errors
(see Sect.~\ref{Sect4-1} for more details).
}
\begin{ruledtabular}
\label{tbl-1}
\begin{tabular}{llcc}
\noalign{\smallskip}
\multicolumn{1}{c}{Transition} & 
\multicolumn{1}{c}{$\nu_{\rm rest}$,} & $\lambda_{\rm rest},$ & 
\multicolumn{1}{c}{$\varepsilon_v$,} \\
 & \multicolumn{1}{c}{GHz} & cm &  \multicolumn{1}{c}{\ms} \\
\noalign{\smallskip}
\hline
\noalign{\medskip}
CCS  $J_N = 2_1-1_0$ & 22.344033(1)\footnotemark[1] & 1.34 & 13.4 \\
NH$_3$   $(J,K) = (1,1)$ & 23.6944955(1)\footnotemark[1] & 1.27 & 1.3  \\
HC$_3$N   $J = 5-4$ & 45.4903102(3)\footnotemark[2] & 0.66 & 2.8  \\
N$_2$H$^+$   $J = 1-0$ & 93.173777(4)\footnotemark[2] & 0.32 & 13.5  \\
\end{tabular}
\end{ruledtabular}
\footnotetext[1]{Rest frequency from Ref.~\onlinecite{RPF08}.} 
\footnotetext[2]{Rest frequency from Ref.~\onlinecite{SSK08}.} 
\end{table}

\section{Interstellar molecules as probes of \dmm\ spatial variations}
\label{Sect2}

Extremely narrow molecular lines observed in cold dark clouds 
provide a sensitive spectroscopic tool to study
the spatial variations of fundamental 
constants under different environments and gravitational fields. 
Among numerous molecules detected in the interstellar medium (ISM),
ammonia (NH$_3$) is of particular interest due to its high
sensitivity to changes in the
electron-to-proton mass ratio, $\mu$.
As was firstly shown for ammonia isotopomer $^{15}$ND$_3$ \cite{VVK04}, 
the inversion frequency in the $(J,K) = (1,1)$
level changes as 
\begin{equation}
\frac{\Delta \nu}{\nu} = 5.6\frac{\Delta \mu}{\mu}\ ,
\label{S2eq1}
\end{equation}
which is nearly an order of magnitude more sensitive to $\mu$-variation
than molecular vibrational frequencies which scale as 
$E_{\rm vib} \sim \mu^{1/2}$. 

The deuterated ammonia shows, however, low abundances in the ISM molecular
clouds, $n({\rm ND}_3)/n({\rm NH}_3) \simeq$ $10^{-4}- 10^{-2}$ \cite{FPF06}.
For more abundant ammonia form of NH$_3$, the sensitivity coefficient of 
the inversion transition $\nu = 23.69$ GHz 
was calculated in \cite{FK07b}: 
\begin{equation}
\frac{\Delta \nu}{\nu} = 4.46\frac{\Delta \mu}{\mu}\ .
\label{S2eq2}
\end{equation}
By comparing the observed inversion frequency of NH$_3$ (1,1) with a suitable 
rotational frequency (scaling as $E_{\rm rot} \sim \mu$)
of another molecule arising {\it co-spatially} with ammonia, 
a limit on the spatial variation of $\mu$ can be determined \cite{FK07b}:
\begin{equation}
\frac{\Delta \mu}{\mu} = 0.289\frac{V_{\rm rot} - V_{\rm inv}}{c}
\equiv 0.289\frac{\Delta V}{c}\ ,
\label{S2eq3}
\end{equation}
where $V_{\rm rot}$ and $V_{\rm inv}$ are the apparent radial velocities
of the rotational and inversion transitions, respectively.

In the present paper we use the comparison of 
the relative radial velocities of ammonia inversion lines 
and rotational transitions of another N-bearing
(N$_2$H$^+$) and C-bearing (CCS, HC$_3$N)
molecules (Table~\ref{tbl-1})
to set a limit on spatial variations of \dmm.
Here we consider some physical and kinematic properties of the
molecular gas clouds where these molecules are detected.

%-----------------Figure 1
%\begin{figure}[htbp!]
\begin{figure}
\includegraphics[viewport=19 167 570 720,width=80mm,height=60mm]{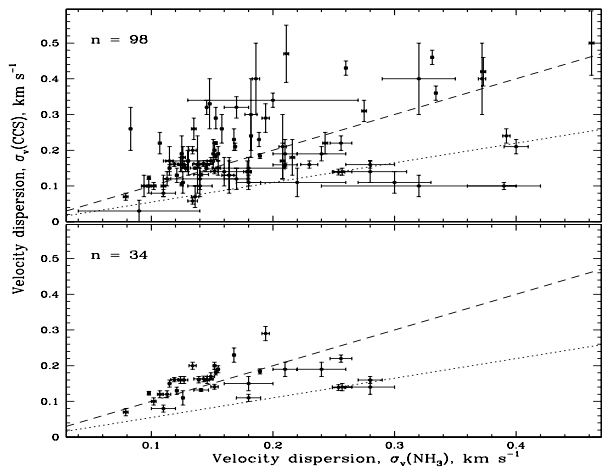}
\caption{\label{fg1} 
{\it Upper panel}: CCS ($2_1-1_0$) versus NH$_3$(1,1) linewidths
for cores in the Perseus molecular cloud from the total sample 
of Ref.~\cite{RPF08}.
The error bars represent $1\sigma$ standard
deviations. {\it Lower panel}: Subsample of the best quality single-component
profiles of CCS and NH$_3$ selected from the full set of
the observed spectra available online (see Table~\ref{tbl-2}).
In both panels, the dashed and dotted lines are 
the boundaries for,
respectively, pure turbulent, $\sigma_v({\rm CCS}) = \sigma_v({{\rm NH}_3})$,
and pure thermal, $\sigma_v({\rm CCS}) = 0.55 \sigma_v({{\rm NH}_3})$,
line broadening. The sample sizes are indicated in the panels.
}
\end{figure}

NH$_3$, N$_2$H$^+$, CCS, and HC$_3$N 
are usually observed in emission in dense molecular clouds
($n_{\scriptscriptstyle \rm H} \ga 10^4$ \cmm) which 
are represented by a large variety of types
in the Milky Way disk \cite{DFE07}. 
Two main classes are high-mass clumps (${\cal M} \ga 100{\cal M}_\odot$) 
associated with infrared dark clouds (IRDCs), and low-mass cores 
(${\cal M} \la 10{\cal M}_\odot$) 
subdivided in turn into protostellar cores
containing infrared sources, and starless cores (pre-protostellar, or
prestellar cores) without embedded luminous sources of any mass. 
The latter provide information on the physical conditions of dense molecular
gas preceding gravitational collapse. 
Depending on the critical gas density in the core center 
($n_{{\scriptscriptstyle \rm H}_2} \sim 10^5$ \cmm), the starless cores can be
either dynamically stable against gravitational contraction 
($n_{{\scriptscriptstyle \rm H}_2} \la 10^5$ \cmm), or unstable 
if density exceeds this critical value \cite{KC08}.
The dense cores reside within a
larger molecular complex or can be isolated. With increasing angular resolution
the cores reveal a substructure with its own range of masses. 
Being very sensitive to the environmental physical conditions,
different molecules probe, in general, different substructures
within the clouds. 

%-----------------Figure 2
\begin{figure}[htbp!]
\includegraphics[viewport=19 167 570 720,width=80mm,height=60mm]{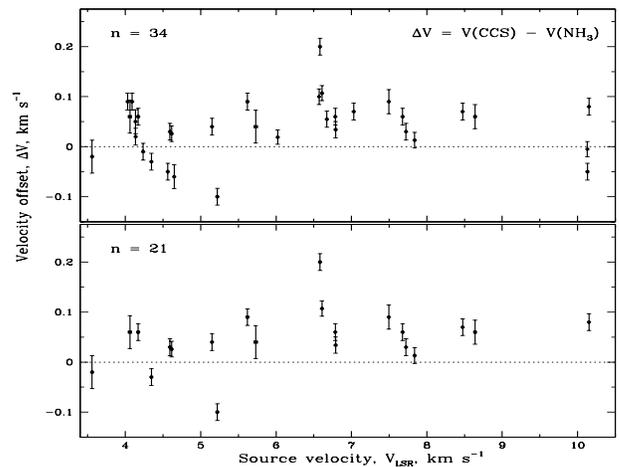}
\caption{\label{fg2}
{\it Upper panel}: 
Velocity offset $\Delta V_{{\rm CCS-NH}_3}$ 
versus the source radial velocity 
for points shown in Fig.~\ref{fg1}, lower panel.
{\it Lower panel}: Same as the upper panel but for the points
marked by asterisks in Table~\ref{tbl-2}. 
The vertical error bars include both random and systematic errors caused by
the uncertainties of the adopted rest frequencies (see Table~\ref{tbl-1}).
}
\end{figure}

At high spectral resolution available in the microwave range 
the errors in the molecular line position measurements 
are mainly restricted by the uncertainties
in laboratory frequencies, $\varepsilon_\nu \sim 0.1-1$ kHz, 
which correspond to the $V_{\rm LSR}$ uncertainties of 
$\varepsilon_v \sim 1-10$ \ms\ \cite{RPF08}.
Taking into account that \dmm $\sim 0.3\Delta V/c$,
the expected precision of the \dmm\ measurements in the ISM is 
restricted by the level of $\sim 10^{-8}$, i.e. 
they are about 100 times more accurate 
than the \dmm\ estimate deduced at $z = 0.68$ \cite{FK07b}.

This level of accuracy
can be achieved, however, only ideally since it assumes that 
molecules are identically (co-spatially)
distributed within the cloud, and are observed
simultaneously with the same receiver, beam size, system temperature, and
velocity resolution.
Violation of any of these conditions leads to
shifts of the line centers which are 
referred to as the {\it Doppler noise}.
The input of the Doppler noise to a putative \dmm\ signal can 
be reduced to some extent if the velocity shifts due to
inhomogeneous distribution of molecules and to 
instrumental imperfections are of random nature.

All molecular transitions listed in Table~\ref{tbl-1}
require for their excitation high gas density,  
$n_{\scriptscriptstyle \rm H} \sim 10^4$ \cmm,  at
a typical kinetic temperature of $T_{\rm kin} \sim 10$ K.
Ammonia is easily detectable in many interstellar clouds. 
The inversion transition $(J,K) = (1,1)$ has 18 hyperfine components
which can be resolved into nine separate features for quiescent nearby
dark clouds \cite{HT83}. 
Since hyperfine components have different
intensities, a single NH$_3$ spectrum allows to determine optical depth,
radial velocity, intrinsic linewidth, and excitation temperature and thus
to separate the effects of optical depth and excitation temperature \cite{BM89}.

%-------------------------- Table 2
\begin{table*}[tbh]
\caption{Parameters of dense cores in the Perseus molecular 
cloud ($D \simeq 260$ pc)
taken from Ref.~\cite{RPF08} and representing the best quality single
component profiles of CCS ($2_1-1_0$) and NH$_3$ (1,1).  
Asterisk marks objects with\, 
$0.55\sigma_v({{\rm NH}_3}) \la \sigma_v({\rm CCS}) \la \sigma_v({{\rm NH}_3})$ 
(see Fig.~\ref{fg1}). 
}
\begin{ruledtabular}
\label{tbl-2}
\begin{tabular}{lldddd}
\noalign{\smallskip}
NH3SRC & Other & \multicolumn{1}{c}{$V_{\rm LSR}$} & 
\multicolumn{1}{c}{$\sigma_v({{\rm NH}_3})$} &
\multicolumn{1}{c}{$\sigma_v({\rm CCS})$} & 
\multicolumn{1}{c}{$\Delta V_{{\rm CCS-NH}_3}$} \\ 
source & name & \multicolumn{1}{c}{(\kms)} & 
\multicolumn{1}{c}{(\kms)} & \multicolumn{1}{c}{(\kms)} & 
\multicolumn{1}{c}{(\kms)} \\
\noalign{\smallskip}
\hline
\noalign{\medskip}
 3 & B16 & 4.136(1) & 0.115(1) & 0.15(1) & 0.02(1)\\[-2pt]
 6 & B11 & 4.030(3) & 0.134(3) & 0.20(1) & 0.09(1)\\[-2pt]
 7 & B30 & 4.135(1) & 0.168(1) & 0.23(2) & 0.05(2)\\[-2pt]
 8 & S57 & 4.090(1) & 0.155(1) & 0.19(1) & 0.09(1)\\[-2pt]
 9 & B13 & 4.563(2) & 0.124(2) & 0.16(1) &-0.05(1)\\[-2pt]
 11& B6  & 4.236(8) & 0.143(8) & 0.163(8)&-0.01(1)\\[-2pt]
 13$\ast$& B1  & 4.06(1)  & 0.24(2)  & 0.19(1) &0.06(2)\\[-2pt]
 15& B27 & 4.648(3) & 0.194(3) & 0.29(2) &-0.06(2)\\[-2pt]
 16$\ast$ & B4  & 4.614(7) & 0.141(6) & 0.132(4)&0.026(8)\\[-2pt]
 19$\ast$& B22 & 4.345(7) & 0.256(9) & 0.22(1) &-0.03(1)\\[-2pt]
 20$\ast$& B21 & 4.17(1)  & 0.28(1)  & 0.16(1) & 0.06(1)\\[-2pt]
 21$\ast$& B20 & 3.559(9) & 0.21(1)  & 0.19(2) &-0.02(3)\\[-2pt]
 22$\ast$& B10 & 5.150(2) & 0.146(2) & 0.16(1) &0.04(1)\\[-2pt]
 24$\ast$& B9  & 4.591(3) & 0.149(3) & 0.17(1) &0.03(1)\\[-2pt]
 28$\ast$& D7  & 5.62(1)  & 0.11(1)  & 0.08(1) &0.09(1)\\[-2pt]
 30$\ast$& L24 & 5.22(1)  & 0.18(1)  & 0.11(1) &-0.10(1)\\[-2pt]
 46& B45 & 7.030(1) & 0.152(1) & 0.20(1) &0.07(1)\\[-2pt]
 60$\ast$& D4  & 5.73(2)  & 0.18(2)  & 0.15(2) &0.04(3)\\[-2pt]
 87$\ast$& B85 & 7.495(1) & 0.126(1) & 0.11(2) &0.09(2)\\[-2pt]
 93& B18 & 6.022(1) & 0.098(1) & 0.123(5)&0.019(5)\\[-2pt] 
 96$\ast$& B34 & 7.838(3) & 0.152(3) & 0.141(7)&0.013(8)\\[-2pt]
 97$\ast$& B32 & 7.724(2) & 0.102(2) & 0.10(1) &0.03(1)\\[-2pt]
 111&B40 & 6.673(1) & 0.139(1) & 0.161(9)&0.055(9)\\[-2pt]
 112$\ast$&B75 & 6.583(7) & 0.257(8) & 0.14(1) &0.20(1)\\[-2pt]
 113&B41 & 6.570(1) & 0.153(1) & 0.182(8)&0.100(8)\\[-2pt]
 114$\ast$&B42 & 6.609(1) & 0.1894(9)& 0.184(7)&0.107(7)\\[-2pt]
 124$\ast$&B44 & 6.786(1) & 0.121(1) & 0.13(1) &0.06(1)\\[-2pt]
 126$\ast$&B48 & 6.790(4) & 0.254(5) & 0.139(8)&0.034(9)\\[-2pt]
 128$\ast$&B58 & 7.678(2) & 0.079(2) & 0.07(1) &0.06(1)\\[-2pt]
 142$\ast$&B91 & 8.475(3) & 0.113(3) & 0.12(1) &0.07(1)\\[-2pt]
 151$\ast$&BS9 & 8.64(1)  & 0.28(2)  & 0.14(2) &0.06(2)\\[-2pt]
 188$\ast$&B121& 10.151(2)& 0.107(2) & 0.12(1) &0.08(1)\\[-2pt]
 190&L44 & 10.130(3)& 0.119(3) & 0.161(6)&-0.005(7)\\[-2pt]
 191&D28 & 10.131(3)& 0.127(3) & 0.16(1) &-0.05(1)\\
\end{tabular}
\end{ruledtabular}
\end{table*}

The second N-bearing molecule, which is 
a good tracer of cold and quiescent dense gas, is the
molecular ion N$_2$H$^+$.
The seven hyperfine components of the N$_2$H$^+$ $J=1-0$
transition were studied with high
spectral resolution in \cite{CMT95}. 
N$_2$H$^+$ is mainly formed in well-shielded regions where
photoprocesses are unimportant \cite{Tu95}.
The intensity distribution of N$_2$H$^+$ $(1-0)$ in dense cores 
closely matches that of NH$_3$ (1,1), and the relative abundances of these
molecules are almost constant across the starless core. However, in
some cases there were detections of ammonia abundances increased by a factor of
$\sim$1-10 towards the core center \cite{TMC04}.
High angular resolution observations of nearby prestellar objects show that
N-bearing molecules trace the inner cores where the density approaches
$10^5$ \cmm\ and the carbon-chain molecules disappear from the gas-phase
because of the freeze-out onto dust grains \cite{DFE07}.
The observed correlation between N$_2$H$^+$
and NH$_3$ velocities and linewidths is another indication 
of probable co-spatial origin of these two species \cite{CBM02}.

%-----------------Figure 3
\begin{figure}[htbp!]
\centering
\includegraphics[viewport=50 167 600 720,width=80mm,height=60mm]{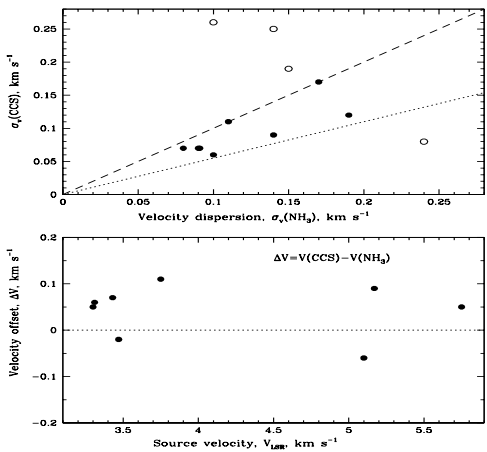}
\caption{\label{fg3}
{\it Upper panel}: 
CCS ($2_1-1_0$) versus NH$_3$(1,1) linewidths
for dense cores associated with the Pipe Nebulae 
from the survey of Ref.~\cite{RLM08}.
No error bars were reported in the original paper.
The filled circles represent points marked by asterisks in Table~\ref{tbl-3}.
The dashed and dotted lines are the boundaries for,
respectively, pure turbulent, $\sigma_v({\rm CCS}) = \sigma_v({{\rm NH}_3})$,
and pure thermal, $\sigma_v({\rm CCS}) = 0.55 \sigma_v({{\rm NH}_3})$,
line broadening. 
{\it Lower panel}:
Velocity offset $\Delta V_{{\rm CCS}-{\rm NH}_3}$ versus the source radial 
velocity set by $V_{\rm LSR}$(NH$_3$).
}
\end{figure}

The spectral lines of the radical CCS without hyperfine splittings and not
strongly saturated ($\tau < 3$) make CCS a well-suited molecule for
detailed studies of the velocity structure in dark clouds \cite{SKY87}.
CCS lines are rather intense in starless, cold and quiescent dark clouds
at the earliest stages of molecular cloud evolution, 
whereas NH$_3$, as mentioned above, tends to be abundant in more chemically
evolved and dense central regions of molecular cores \cite{SYO92}.
As a rule, CCS shows up as a 
clumpy distributed gas slightly outside these central regions.
A pronounced spatial {\it anticorrelation} in the emission from CCS and NH$_3$
was observed on large scale 
maps of the Taurus Molecular Cloud 1 (TMC-1) ridge \cite{HSY92},
pre-protostellar cores L1498 \cite{KLV96} 
and B68 \cite{LVL03}, as well as on small scale maps with angular
resolution of $\simeq 5''$ (= 1750 AU) in the infrared cloud B1-IRS 
where the CCS emission was detected from three clumps surrounding
the central source \cite{DGM05}.  The clumps in B1-IRS
exhibit a velocity gradient from red- to blueshifted velocities 
with respect to the systemic velocity of the central source which can be
interpreted as interaction of the CCS-emitting gas with the molecular outflow. 
This interaction makes CCS lines wider than their thermal widths
and comparable with the measured widths of the NH$_3$ lines in the B1-IRS
source: $\sigma_v({\rm CCS}) = 0.9\pm0.1$ \kms\ and 
$\sigma_v({\rm NH}_3) = 1.1\pm0.1$ \kms
(CCS thermal linewidth is 
narrower by a factor of 0.55 than that of NH$_3$ due to the different masses
of these molecules). Turbulent broadening could equalize the
apparent linewidths if both CCS and  NH$_3$ trace the same volume elements.
However, in 5 out of 6 low-mass young 
star-forming regions studied in \cite{DGM06}
the widths of the CCS lines are too narrow,
$\sigma_v({\rm CCS}) < 0.55 \sigma_v({\rm NH}_3)$, 
which is consistent with both molecules tracing different
regions of gas with different kinematics. 
The only exception in this sample
is the B1-IRS cloud where the spatial anticorrelation in 
the CCS- and NH$_3$-emitting gas has been revealed \cite{DGM05}.
On the other hand, from the survey of dense cores using lines of
N$_2$H$^+$, C$_3$H$_2$, and CCS it was deduced \cite{BCM98}
that velocities and linewidths of these three molecules are well correlated
among themselves and with NH$_3$, 
indicating a kinematic consistency for these species.

The three strongest hyperfine components of the $J = 5-4$ 
rotational transition of HC$_3$N
\cite{LL78} were partly resolved in radio-astronomical observations \cite{YSK90}.
HC$_3$N, as well as CCS, is abundant in the early stage
of chemical evolution of the star-forming regions \cite{LHP96} when
the ratio of the fractional abundances of HC$_3$N and CCS remains almost
constant \cite{SYO92} and the spatial distributions of these C-bearing
molecules match each other quite well.
The continued chemical evolution leads, however, to
the spatial variation of the relative molecular abundances 
(chemical differentiation)   
due to adsorption of the heavy elements from the gas phase 
onto grain mantles at high densities and dust temperatures 
$T_{\rm D} < 20$ K \cite{FPF06}. 
In the later stages of the protostellar collapse carbon 
chain molecules are destroyed 
by high velocity outflows and radiation from protostars, whereas the same
processes favor the desorption of ammonia from dust grains \cite{SYO92}.

Thus, the spatial distribution of the emitting gas tracers 
in star-forming regions is governed by 
both the physical conditions
within the dense cores and the time-dependent chemistry.
The consistency of the apparent linewidths of different molecules
is a necessary condition for these molecules
to be co-spatially distributed, but is not a sufficient one.  

Molecular line profiles are also affected by the bulk kinematics 
of the dense clouds. 
Plane-of-sky bulk motions observed in starless cores exhibit 
velocity gradients of 
$\sim$0.3 to 1.4 km s$^{-1}$ pc$^{-1}$ \cite{BG98,SWD05}.
A differential rotation and angular momentum evolution in the B68 core with
velocity gradients of 3.4 and 4.8 km s$^{-1}$ pc$^{-1}$ for
the outer and inner material respectively was found in \cite{LBA03}.
Similarly, the spatial variations of the velocity gradients in 
prestellar and protostellar cores of $\sim$2 km s$^{-1}$ pc$^{-1}$ \cite{CBM02}
indicate deviations from simple solid body rotation. 

The gas kinematics along the line-of-sights studied in the isolated cores 
\cite{WMB04} shows that cores move 
slowly ($\la 0.1$ \kms) relatively to their surroundings.  
Inward and/or outward motion with velocity of $\sim$0.1 \kms\ 
is revealed by the skewed line profiles. 
For instance, blue asymmetry of the 
hyperfine components of the  N$_2$H$^+$
(1-0) transition seen in two starless cores can be produced by absorption
along the line-of-sight in the foreground inward-moving gas of lower
excitation \cite{WLM06}.
Acoustic type oscillations with amplitude of $\sim$0.1 \kms\  are seen in
the inward and outward gas motion across the B68 core \cite{LBA03}.
A redshifted peak of the CCS $3_2-2_1$ emission line 
of $\sim$0.16 \kms\ with respect to the central velocity of the NH$_3$ (1,1) line
was detected in the L1551 dark cloud \cite{SWD05}. 

The turbulent component of the gas velocity derived in \cite{TMC04}
from the intrinsic linewidths of the NH$_3$ and N$_2$H$^+$ lines  
is small and subsonic, $\sigma_{\rm tur} \la 0.1$ \kms\ 
(for $T_{\rm kin} = 10$ K, sound speed is 0.19 \kms).
Earlier, a similar conclusion has been obtained from
NH$_3$ observations of low-mass dense cores \cite{M83}.

%-------------------------- Table 3
\begin{table}[t!]
%\centering
\caption{Parameters of dense cores in the Pipe Nebula ($D \simeq 130$ pc)
taken from Ref.~\cite{RLM08} and representing emission line
profiles of CCS ($2_1-1_0$) and NH$_3$ (1,1). 
Asterisk marks filled circles in Fig.~\ref{fg3}.
}
\begin{ruledtabular}
\label{tbl-3}
\begin{tabular}{lddddd}
\noalign{\smallskip}
\multicolumn{1}{c}{Source} & \multicolumn{2}{c}{$V_{\rm LSR}$, \kms} & 
\multicolumn{2}{c}{$\sigma_v$, \kms} & \multicolumn{1}{c}{Mass,} \\
\multicolumn{1}{c}{\#} & \multicolumn{1}{c}{CCS} & \multicolumn{1}{c}{NH$_3$} & 
\multicolumn{1}{c}{CCS} & \multicolumn{1}{c}{NH$_3$} & 
\multicolumn{1}{c}{${\cal M}_\odot$} \\
\noalign{\smallskip}
\hline
\noalign{\medskip}
6$\ast$   &3.50 &3.43 &0.07  &0.09 &3.1\\[-2pt] 
12        &3.47 &3.20 &0.19  &0.15 &20.4\\[-2pt]
14$\ast$  &3.45 &3.47 &0.09  &0.14 &9.7\\[-2pt]  
40$\ast$  &3.35 &3.30 &0.06  &0.10 &9.2\\[-2pt]  
42$\ast$  &3.86 &3.75 &0.11  &0.11 &2.8\\[-2pt]  
87        &4.50 &4.47 &0.25  &0.14 &10.3\\[-2pt]
89        &4.87 &4.45 &0.26  &0.10 &1.4\\[-2pt]
92$\ast$  &5.04 &5.10 &0.12  &0.19 &1.6\\[-2pt]  
93$\ast$  &5.26 &5.17 &0.17  &0.17 &3.5\\[-2pt]
101$\ast$ &3.37 &3.31 &0.07  &0.09 &1.9\\[-2pt]
102       &4.84 &4.87 &0.08  &0.24 &6.7\\[-2pt]
109$\ast$ &5.80 &5.75 &0.07  &0.08 &3.6\\
\end{tabular}
\end{ruledtabular}
\end{table}

%-----------------Figure 4
\begin{figure}[htbp!]
\includegraphics[viewport=19 167 570 720,width=80mm,height=60mm]{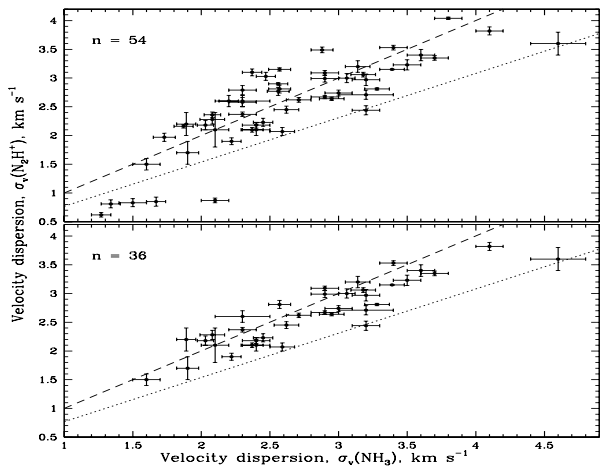}
\caption{\label{fg4}
{\it Upper panel}: 
N$_2$H$^+$ ($1-0$) versus NH$_3$(1,1) linewidths
for massive clumps associated with infrared dark clouds from the survey of
Ref.~\cite{SSK08}. The error bars represent $1\sigma$ standard deviations. 
{\it Lower panel}: 
Same as the upper panel but for the points
marked by asterisks in Table~\ref{tbl-4}.
In both panels, the dashed and dotted lines are 
the boundaries for,
respectively, pure turbulent, $\sigma_v({\rm N}_2{\rm H}^+) = 
\sigma_v({{\rm NH}_3})$,
and pure thermal, $\sigma_v({\rm N}_2{\rm H}^+) = 0.77 \sigma_v({{\rm NH}_3})$,
line broadening. The sample sizes are indicated in the panels.
}
\end{figure}

The velocity widths in high-mass clumps associated with IRDCs
are large ($\sim$1-3 \kms) and indicate the supersonic motion.
If the IRDCs are high-mass star-forming regions, then,
as noted in Ref.~\cite{SSK08}, 
the observed large velocity widths are consistent with a model which suggests
massive star formation in turbulence-supported cores with a high accretion rate
\cite{MT03}.
The innermost regions of such clumps are traced by molecular
transitions with relatively large excitation energies of the upper
states. To excite such transitions denser and hotter regions closer
to the protostar are required. 
The NH$_3$ (2,2) and (3,3) transitions have, for example,
the excitation energies of 
$E^{\scriptscriptstyle(2,2)}_{\rm u} = 65$ K and
$E^{\scriptscriptstyle(3,3)}_{\rm u} = 125$ K, 
and so their velocity distributions
should be broader than that of the
NH$_3$ (1,1) line for which 
$E^{\scriptscriptstyle(1,1)}_{\rm u} = 23$ K. 
We may expect that other molecules, such as 
CCS ($J_N = 2_1-1_0$), HC$_3$N ($J = 5-4$), and N$_2$H$^+$ ($J=1-0$)
with excitation energies lower than 23 K,
would have velocity distributions 
similar to that of NH$_3$ (1,1).

%-----------------Figure 5
\begin{figure}[htbp!]
\includegraphics[viewport=19 167 570 720,width=80mm,height=60mm]{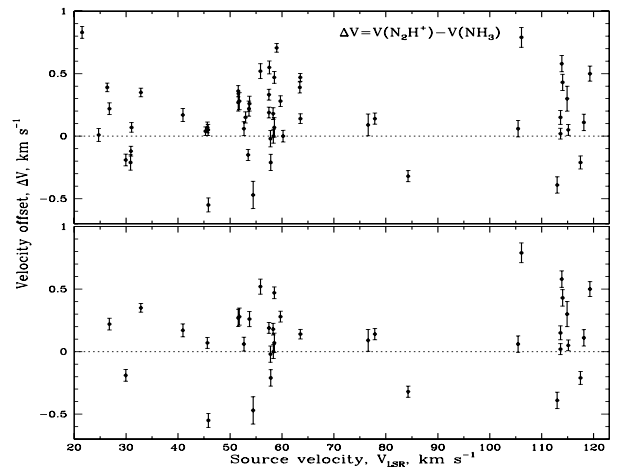}
\caption{\label{fg5}
{\it Upper panel}: 
Velocity offset $\Delta V_{{\rm N}_2{\rm H}^+-{\rm NH}_3}$ 
versus the source radial 
velocity set by $V_{\rm LSR}$(NH$_3$). 
{\it Lower panel}: Same as the upper panel but for the
points marked by asterisks in Table~\ref{tbl-4}. 
The vertical error bars include both random and systematic errors caused by
the uncertainties of the adopted rest frequencies (Table~\ref{tbl-1}).
}
\end{figure}

\section{Results}
\label{Sect3}

As noted in the previous section, the observational constraint on \dmm\
is set by the velocity offset, $\Delta V$, measured between
the ammonia inversion 
line and a molecular rotational line, Eq.(\ref{S2eq3}). 
In this comparison, to minimize possible systematics in the $\Delta V$ values,
the second molecule must follow closely the spatial distribution of NH$_3$.  
The most favorable candidate which fits this condition is the molecular
ion N$_2$H$^+$. As for carbon-chain molecules, it seems likely that 
CCS and HC$_3$N are anticorrelated with NH$_3$, albeit not all of the dense
cores display chemical differentiation (Ref.~\onlinecite{DFE07}, Sect. 4.4). 
When molecules trace the same gas, their turbulent component should be equal,
and the lighter NH$_3$ should always have larger linewidth if 
turbulent and thermal line broadening are comparable,
$\sigma_{\rm tur}/\sigma_{\rm therm} \sim 1$. If, however, 
the turbulent component dominates, the lines may have similar widths.
We use this criterion to select molecular pairs tracing presumably
the same gas velocity, i.e. arising co-spatially. 

It is to note that the velocity offset $\Delta V$  in Eq.(\ref{S2eq3}) is
the sum of two components, $\Delta V = \Delta V_\mu + \Delta V_n$, with
$\Delta V_\mu$ being the shift due to $\mu$-variation, and $\Delta V_n$ 
the shift induced by Doppler noise.
Assuming the random character of the Doppler noise,
the signal $\Delta V_\mu$ can be estimated statistically by
averaging over a large data sample. 
The validity of this assumption requires, however, 
additional tests as discussed below in
Sect.~\ref{Sect3-3}.

%-------------------------- Table 4
\begin{table*}[t!]
\caption{Parameters of massive clumps associated with infrared dark clouds
taken from Ref.~\cite{SSK08} and representing emission lines of
N$_2$H$^+$ ($1-0$) and NH$_3$ (1,1). 
Asterisk marks objects with\, 
$0.77\sigma_v({{\rm NH}_3}) \la \sigma_v({\rm N}_2{\rm H}^+) \la \sigma_v({{\rm NH}_3})$
(see Fig.~\ref{fg4}, lower panel). 
}
\begin{ruledtabular}
\label{tbl-4}
\begin{tabular}{lddddd}
\noalign{\smallskip}
\multicolumn{1}{c}{Source} & \multicolumn{2}{c}{$V_{\rm LSR}$, \kms} &  
\multicolumn{2}{c}{$\sigma_v$, \kms} & $D$, \\
 & \multicolumn{1}{c}{N$_2$H$^+$} & \multicolumn{1}{c}{NH$_3$} &  
\multicolumn{1}{c}{N$_2$H$^+$} & \multicolumn{1}{c}{NH$_3$} & kpc \\
\noalign{\smallskip}
\hline
\noalign{\medskip}
G015.05+00.07 MM1   &24.73(3) &24.72(4) &2.79(8) &2.3(1)  &2.5\\[-2pt]
G015.31--00.16 MM2  &31.12(2) &31.05(3) &0.81(7) &1.34(7) &3.1\\[-2pt]
G015.31--00.16 MM3  &30.81(2) &30.93(3) &0.62(4) &1.27(7) &3.0\\[-2pt]
G019.27+00.07 MM1   &26.76(1) &26.37(3) &3.15(3) &2.57(8) &2.3\\[-2pt]
G019.27+00.07 MM2$\ast$   &27.01(2) &26.79(4) &2.62(4) &2.71(9) &2.3\\[-2pt]
G022.35+00.41 MM1$\ast$   &52.73(2) &52.67(5) &2.74(5) &3.0(1)  &3.7\\[-2pt]
G022.35+00.41 MM2   &60.22(2) &60.22(4) &0.83(7) &1.5(1)  &4.1\\[-2pt]
G022.35+00.41 MM3$\ast$   &83.97(3) &84.29(3) &2.23(7) &2.45(7) &5.3\\[-2pt]
G023.60+00.00 MM1$\ast$   &106.89(6)&106.10(5) &3.82(7) &4.1(1)  &6.7\\[-2pt]
G023.60+00.00 MM2   &53.34(3) &53.49(3) &3.03(7) &2.47(7) &3.7\\[-2pt]
G023.60+00.00 MM3$\ast$   &105.51(4)&105.45(5) &3.4(1)  &3.6(1)  &6.6\\[-2pt]
G023.60+00.00 MM4   &53.87(5) &53.65(3) &2.6(1)  &2.20(7) &3.7\\[-2pt]
G023.60+00.00 MM7$\ast$   &54.00(5) &53.74(3) &1.90(6) &2.22(7) &3.7\\[-2pt]
G024.08+00.04 MM1$\ast$   &113.63(1)&113.61(4) &2.67(3) &2.9(1)  &7.8\\[-2pt]
G024.08+00.04 MM2(1)$\ast$&52.08(3) &51.80(6) &1.5(1)  &1.6(1)  &3.6\\[-2pt]
G024.08+00.04 MM2(2)$\ast$&114.45(4)&114.02(5) &2.1(1)  &2.4(1)  &7.8\\[-2pt]
G024.08+00.04 MM3$\ast$   &51.86(6) &51.59(3) &2.2(2)  &1.89(7) &3.6\\[-2pt]
G024.08+00.04 MM4   &51.95(3) &51.59(3) &0.85(8) &1.67(7) &3.6\\[-2pt]
G024.33+00.11 MM1$\ast$   &113.76(2)&113.61(5) &3.53(4) &3.4(1)  &7.7\\[-2pt]
G024.33+00.11 MM2$\ast$   &118.22(4)&118.11(5) &3.23(9) &3.5(1)  &7.7\\[-2pt]
G024.33+00.11 MM4$\ast$   &115.19(7)&114.89(7) &3.6(2)  &4.6(2)  &7.7\\[-2pt]
G024.33+00.11 MM5$\ast$   &117.26(4)&117.47(3) &3.0(8)  &3.06(6) &7.7\\[-2pt]
G024.33+00.11 MM6$\ast$   &114.43(4)&113.85(5) &2.97(10)&3.2(1)  &7.7\\[-2pt]
G024.33+00.11 MM9$\ast$   &119.78(5)&119.28(3) &1.7(2)  &1.90(8) &7.7\\[-2pt]
G024.33+00.11 MM11$\ast$  &112.60(5)&112.99(4) &2.18(5) &2.4(1)  &7.7\\[-2pt]
G024.60+00.08 MM1   &53.14(3) &52.99(3)  &3.10(6) &2.37(7) &3.6\\[-2pt]
G024.60+00.08 MM2$\ast$   &115.16(1)&115.11(4) &2.37(4) &2.3(1)  &7.7\\[-2pt]
G025.04--00.20 MM1  &64.00(2) &63.53(2)  &2.36(5) &2.08(6) &4.2\\[-2pt]
G025.04--00.20 MM2  &63.83(3) &63.44(3)  &2.77(7) &2.56(8) &4.3\\[-2pt]
G025.04--00.20 MM4$\ast$  &63.69(2) &63.55(3)  &2.18(8) &2.03(6) &4.1\\[-2pt]
G034.43+00.24 MM1$\ast$   &57.77(5) &57.79(4)  &3.15(1) &3.39(9) &3.5\\[-2pt]
G034.43+00.24 MM2   &57.832(9)&57.50(4)  &4.04(2) &3.8(1)  &3.5\\[-2pt]
G034.43+00.24 MM3   &59.686(8)&58.98(3)  &2.90(2) &2.56(7) &3.6\\[-2pt]
G034.43+00.24 MM4$\ast$   &57.689(8)&57.50(4)  &2.81(2) &3.28(9) &3.5\\[-2pt]
G034.43+00.24 MM5   &58.12(3) &57.57(4)  &2.58(8) &2.3(1)  &3.5\\[-2pt]
G034.43+00.24 MM6$\ast$   &58.49(2) &58.31(4)  &2.07(7) &2.59(9) &3.6\\[-2pt]
G034.43+00.24 MM8$\ast$   &57.63(5) &57.84(4)  &3.2(1)  &3.14(9) &3.5\\[-2pt]
G034.43+00.24 MM9$\ast$   &59.00(2) &58.53(4)  &2.45(6) &2.62(9) &3.6\\[-2pt]
I18102--1800 MM1    &22.36(2) &21.53(4)  &3.49(5) &2.88(8) &2.7\\[-2pt]
I18151--1208 MM1$\ast$    &33.19(1) &32.84(3)  &2.10(4) &2.37(8) &2.8\\[-2pt]
I18151--1208 MM2$\ast$    &29.74(2) &29.93(4)  &3.09(4) &2.9(1)  &2.6\\[-2pt]
I18151--1208 MM3    &30.65(1) &30.86(6)  &0.87(4) &2.1(1)  &2.7\\[-2pt]
I18182--1433 MM1$\ast$    &60.00(1) &59.72(4) & 3.06(4) &3.18(9) &4.6\\[-2pt]
I18182--1433 MM2$\ast$    &41.10(3) &40.93(4)  &2.28(8) &2.08(9) &3.6\\[-2pt]
I18223--1243 MM1    &45.32(1) &45.28(3)  &2.16(3) &1.87(7) &3.6\\[-2pt]
I18223--1243 MM2$\ast$    &45.29(2) &45.84(5)  &2.99(6) &2.9(1)  &3.6\\[-2pt]
I18223--1243 MM3$\ast$    &45.73(1) &45.66(4)  &2.64(3) &2.95(9) &3.7\\[-2pt]
I18223--1243 MM4    &45.83(3) &45.78(3)  &1.97(7) &1.73(8) &3.7\\[-2pt]
I18306--0835 MM1$\ast$    &78.02(3) &77.88(3)  &2.81(7) &2.57(8) &4.9\\[-2pt]
I18306--0835 MM2$\ast$    &76.67(5) &76.58(7)  &2.6(1)  &2.3(2)  &4.9\\[-2pt]
I18306--0835 MM3$\ast$    &53.98(9) &54.45(6)  &2.1(3)  &2.1(1)  &3.7\\[-2pt]
I18337--0743 MM1$\ast$    &58.37(2) &58.37(5)  &3.35(4) &3.7(1)  &3.9\\[-2pt]
I18227--0743 MM2$\ast$    &58.60(3) &58.53(7)  &2.71(8) &3.2(2)  &3.9\\[-2pt]
I18227--0743 MM3$\ast$    &56.37(3) &55.85(5)  &2.44(8) &3.2(1)  &3.8\\
\end{tabular}
\end{ruledtabular}
\end{table*}

\subsection{Dense cores in the Perseus molecular cloud}
\label{Sect3-1}

At first we consider an ammonia spectral atlas of 193 dense 
protostellar and prestellar cores of low masses in the
Perseus molecular cloud \cite{RPF08}. The spectral observations
of the cores in NH$_3$ (1,1), (2,2),
CCS ($2_1-1_0$) and CC$^{34}$S ($2_1-1_0$) lines were carried out
simultaneously using the 100-m Robert F. Byrd Green Bank Telescope (GBT).
Each target was observed in a single-pointing, frequency-switched mode allowing
the entire NH$_3$ (1,1) complex to remain within the spectral window.
The GBT beam size at 23 GHz is $FWHM = 31''$ or 0.04 pc at the Perseus cloud 
distance ($D \sim 260$ pc). 
The resulting spectral resolution was 24 \ms.
Thus, both molecules were observed with similar angular 
and spectral resolutions. 

%-----------------Figure 6
\begin{figure}[t]
\includegraphics[viewport=19 167 570 720,width=80mm,height=60mm]{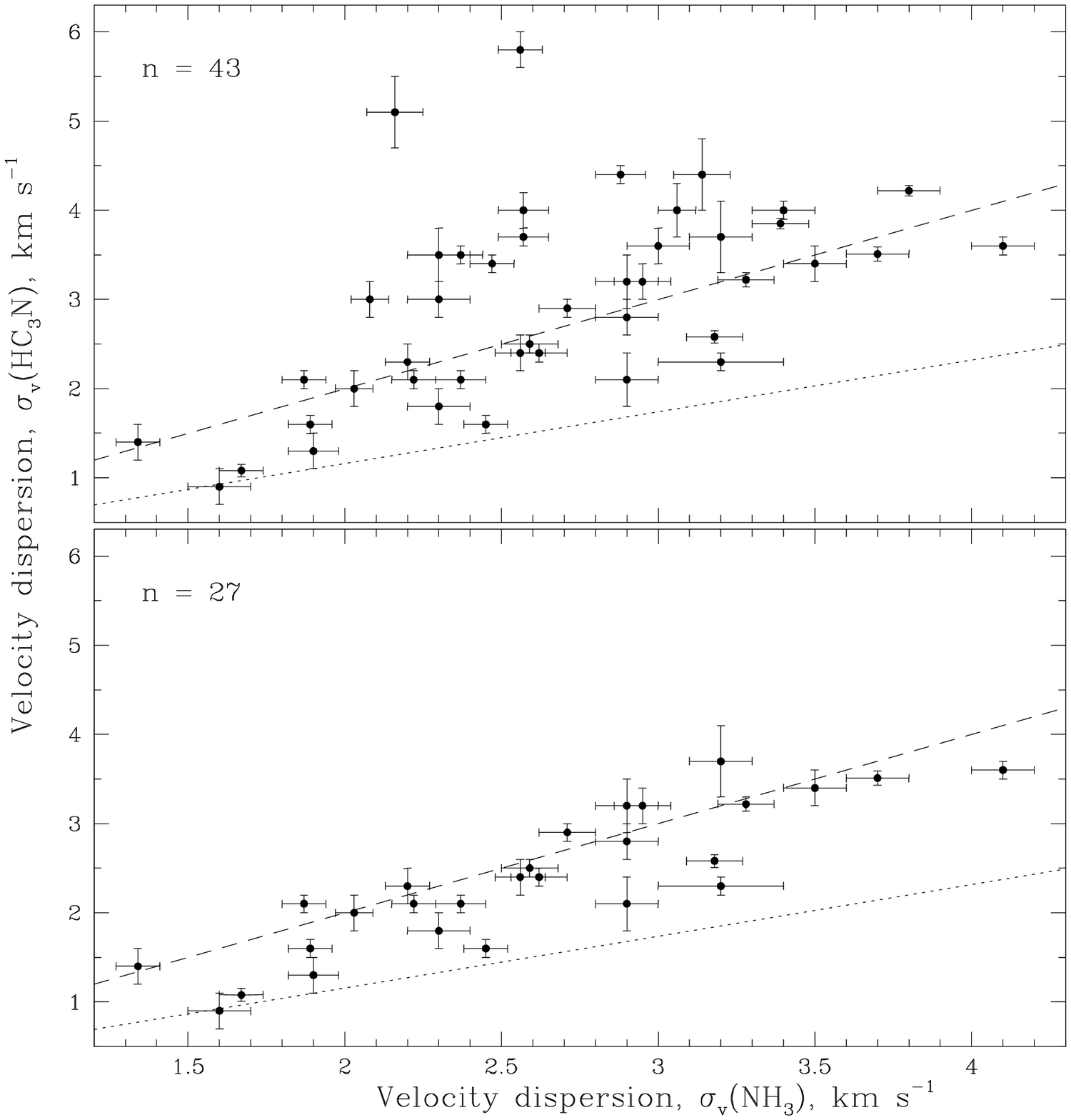}
\caption{\label{fg6}
Same as Fig.~\ref{fg4} but for HC$_3$N ($5-4$) and
NH$_3$(1,1) linewidths.
Points in the lower panel are 
marked by asterisks in Table~\ref{tbl-5}.
In both panels the dashed and dotted lines are 
the boundaries for,
respectively, pure turbulent, $\sigma_v({\rm HC}_3{\rm N}) = 
\sigma_v({{\rm NH}_3})$,
and pure thermal, $\sigma_v({\rm HC}_3{\rm N}) = 0.58 \sigma_v({{\rm NH}_3})$,
line broadening. The sample sizes are indicated in the panels.
}
\end{figure}

The typical cores in Perseus have 
the mean gas density $n_{\scriptscriptstyle \rm H} \sim 
(1-2)\times10^4$ \cmm,
velocity dispersion $\sigma_v = 0.17$ \kms, kinetic temperature
$T_{\rm kin} = 11$ K, radius $R = 0.09$ pc, and mass 
${\cal M} \sim 1\ {\cal M}_\odot$.
The Perseus molecular cloud covers an area of 6 by 2 degrees ($27\times9$ pc)
and contains
over $10^4 {\cal M}_\odot$ of gas and dust that corresponds to the mean gas
density of $n_{\scriptscriptstyle \rm H} \sim 4\times10^2$ \cmm.

The atlas of 193 sources is very conveniently presented on the 
website (see Ref.~\onlinecite{RPF08} for details)
%(http://www.cfa.harvard.edu/COMPLETE/data\_html\_pages/GBT\_NH3.html) 
where the original spectra are shown along with
the best fitting models, model parameters and their uncertainties.
These authors found that many cores show evidence
for the multiple velocity components along the line-of-sight. 
The total number of cores where both 
CCS ($2_1-1_0$) and NH$_3$ (1,1) lines were detected is 98. 
The central velocities of these lines,
being averaged with weights
inversely proportional to the variances of the measurements, reveal 
that CCS lines are systematically
offset by 16 \ms.  
It was suggested in Ref.~\cite{RPF08} that 
this offset is due to uncertainties in the
assumed rest frequency of the CCS line. 
A similar marginally significant offset in the velocities of 
the C$^{18}$O (1-0) and NH$_3$ (1,1) lines was
measured from 18 dark cores in which the same position was observed
in both NH$_3$ (1,1) and C$^{18}$O (1-0)\, \cite{BM89}.

Figure~\ref{fg1}, upper panel, shows linewidths of CCS ($2_1-1_0$) vs.
NH$_3$ (1,1) for a total sample of data points ($n = 98$)
from \cite{RPF08}.
In this panel, the dashed and dotted lines are the boundaries for,
respectively, pure turbulent, $\sigma_v({\rm CCS}) = \sigma_v({{\rm NH}_3})$,
and pure thermal, $\sigma_v({\rm CCS}) = 0.55 \sigma_v({{\rm NH}_3})$,
line broadening. A wide scatter of points reveals the fact that
for many cores a complex structure of line profiles hampers the accurate
estimate of the linewidth. As a next step, we selected from the
database only those profiles which have been perfectly fitted by a single 
component model. The obtained subsample of $n = 34$ pairs is shown
in the lower panel in Fig.~\ref{fg1} with the corresponding fitting
parameters listed in Table~\ref{tbl-2} 
which have been taken from the online database.
The original ammonia data were re-reduced with additional small Doppler
corrections to eliminate IF-to-IF variations (IF stands for Intermediate
Frequency~-- a frequency to which the radio frequency 
is shifted as an intermediate
step before detection in the backend), 
but there were no significant changes from the values 
derived from the original data set \cite{Ro08}.
%(E. W. Rosolowsky, private communication).

Formal calculations of the weighted means for the total sample
and subsample give 
${\Delta V}^w_{n=98} = 0.016\pm0.013$ \kms\ 
(which is in line with Ref.~\onlinecite{RPF08})
and ${\Delta V}^w_{n=34} = 0.040\pm0.010$ \kms, respectively.
However, to cancel out the Doppler noise component $\Delta V_n$,
simple (unweighted) averaging of data is more appropriate
(as seen from Fig.~\ref{fg2}, the $\Delta V$ 
values have approximately equal error bars).
The corresponding unweighted values are
${\Delta V}_{n=98} = 0.044\pm0.013$ \kms\ and
${\Delta V}_{n=34} = 0.039\pm0.010$ \kms.

Now let us select only points with
$\sigma_v({\rm NH}_3) > \sigma_v({\rm CCS}) > 0.55\sigma_v({\rm NH}_3)$,
i.e. points for which both emission lines arise most probably co-spatially
(Fig.~\ref{fg1}, lower panel). This gives $n = 21$ and 
${\Delta V}^w_{n=21} = 0.050\pm0.013$ \kms, and
${\Delta V}_{n=21} = 0.048\pm0.013$ \kms.

The distributions of the selected points are shown in Fig.~\ref{fg2}.
Here the vertical error bars include both the error of the measurement and
the systematic error of $\varepsilon_v = 13.5$ \ms, indicating 
approximate equidispersion of the individual measurements.
It is clearly seen that points with positive offsets dominate.
The scatter of the points reflects complexity of gas kinematics and
effects of chemical segregation of one molecule with respect to the
other in dense molecular cores discussed in Sect.~\ref{Sect2}.

The unweighted mean from the $n = 21$ subsample will be considered
as the final estimate of the velocity offset for the Perseus dark cores.
Accounting for the systematic error due to uncertainties in the
rest frequencies (Table~\ref{tbl-1}), we obtain 
${\Delta V}_{n=21} = 0.048\pm0.013\pm0.0135$ \kms.
The robust redescending $M$-estimate for the shift and scale \cite{HRR86}
yields ${\Delta V}_{n=21} = 0.052\pm0.007\pm0.0135$ \kms\ (see our final
Table~\ref{tbl-6}).

To understand if there are subclasses among the Perseus cores, we also
investigated the measured column densities of NH$_3$ and CCS.
We found the mean ratio of
$N$(NH$_3$)/$N$(CCS) $= 1180\pm870$.
There are only two outlying clouds with 
$N$(NH$_3$)/$N$(CCS) $\sim 5500$ and 25000. 
The plot between the column density ratio versus velocity offset
does not show any trend. 
It seems that the dark cores are rather
homogeneous with respect to the column density ratio, and this
probably means that there are no strong variations in abundances 
among the clouds. In other words there is no evidence for the
two different populations.

%-----------------Figure 7
\begin{figure}[t]
\includegraphics[viewport=19 167 570 720,width=80mm,height=60mm]{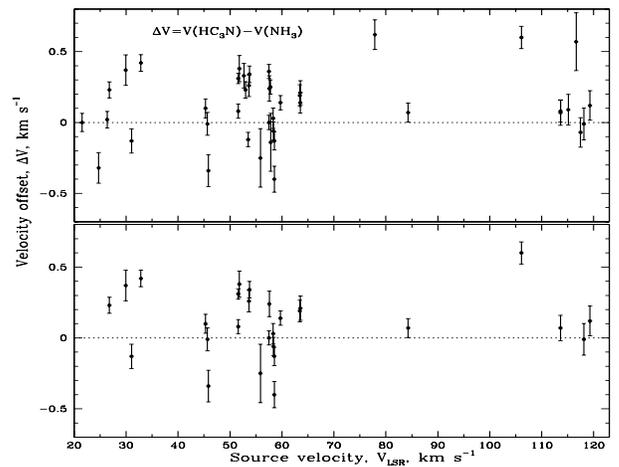}
\caption{\label{fg7}
Same as Fig.~\ref{fg5} but for emission lines 
${\rm HC}_3{\rm N}$ and ${\rm NH}_3$.  
The total sample is shown in Fig.~\ref{fg6}, upper panel.
The points in the lower panel are those marked by asterisks in Table~\ref{tbl-5}. 
}
\end{figure}

%-------------------------- Table 5
\begin{table*}[t!]
\caption{Same as Table~\ref{tbl-4} but for
emission lines of HC$_3$N ($5-4$) and NH$_3$ (1,1). 
Asterisk marks objects with\, 
$0.58\sigma_v({{\rm NH}_3}) \la \sigma_v({\rm HC}_3{\rm N}) \la 
\sigma_v({{\rm NH}_3})$
(see Fig.~\ref{fg6}, lower panel). 
}
\begin{ruledtabular}
\label{tbl-5}
\begin{tabular}{lddddd}
\noalign{\smallskip}
\multicolumn{1}{c}{Source} & \multicolumn{2}{c}{$V_{\rm LSR}$, \kms}  & 
\multicolumn{2}{c}{$\sigma_v$, \kms} & $D$, \\
 & \multicolumn{1}{c}{HC$_3$N} & \multicolumn{1}{c}{NH$_3$}  & 
\multicolumn{1}{c}{HC$_3$N} & \multicolumn{1}{c}{NH$_3$} & kpc \\
\noalign{\smallskip}
\hline
\noalign{\medskip}
G015.05+00.07 MM1   &24.4(1)   &24.72(4)  &3.5(3) &2.3(1)  & 2.5\\[-2pt] 
G015.31--00.16 MM2$\ast$  &30.92(8)  &31.05(3)  &1.4(2) &1.34(7) & 3.1\\[-2pt]
G019.27+00.07 MM1   &26.39(5)  &26.37(3)  &3.7(1) &2.57(8) & 2.3\\[-2pt]
G019.27+00.07 MM2$\ast$   &27.02(4)  &26.79(4)  &2.9(1) &2.71(9) & 2.3\\[-2pt]
G022.35+00.41 MM1   &53.00(7)  &52.67(5)  &3.6(2) &3.0(1)  & 3.7\\[-2pt]
G022.35+00.41 MM3$\ast$   &84.36(6)  &84.29(3)  &1.6(1) &2.45(7) & 5.3\\[-2pt]
G023.60+00.00 MM1$\ast$   &106.70(6) &106.10(5) &3.6(1) &4.1(1)  & 6.7\\[-2pt]
G023.60+00.00 MM2   &53.37(4)  &53.49(3)  &3.4(1) &2.47(7) & 3.7\\[-2pt]
G023.60+00.00 MM4$\ast$   &53.91(7)  &53.65(3)  &2.3(2) &2.20(7) & 3.7\\[-2pt]
G023.60+00.00 MM7$\ast$   &54.08(5)  &53.74(3)  &2.1(1) &2.22(7) & 3.7\\[-2pt]
G024.08+00.04 MM1$\ast$   &113.68(8) &113.61(4) &2.8(2) &2.9(1)  & 7.8\\[-2pt]
G024.08+00.04 MM2(1)$\ast$&52.18(7)  &51.80(6)  &0.9(2) &1.6(1)  & 3.6\\[-2pt]
G024.08+00.04 MM3$\ast$   &51.67(4)  &51.59(3)  &1.6(1) &1.89(7) & 3.6\\[-2pt]
G024.08+00.04 MM4$\ast$   &51.90(2)  &51.59(3)  &1.08(7)&1.67(7) & 3.6\\[-2pt]
G024.33+00.11 MM1   &113.69(6) &113.61(5) &4.0(1) &3.4(1)  & 7.7\\[-2pt]
G024.33+00.11 MM2$\ast$   &118.1(1)  &118.11(5) &3.4(2) &3.5(1)  & 7.7\\[-2pt]
G024.33+00.11 MM3   &117.2(2)  &116.63(4) &5.1(4) &2.16(9) & 7.7\\[-2pt]
G024.33+00.11 MM5   &117.4(1)  &117.47(3) &4.0(3) &3.06(6) & 7.7\\[-2pt]
G024.33+00.11 MM9$\ast$   &119.4(1)  &119.28(3) &1.3(2) &1.90(8) & 7.7\\[-2pt]
G024.60+00.08 MM1   &53.22(5)  &52.99(3)  &3.5(1) &2.37(7) & 3.6\\[-2pt]
G024.60+00.08 MM2   &115.2(1)  &115.11(4) &3.0(2) &2.3(1)  & 7.7\\[-2pt]
G025.04--00.20 MM1  &63.67(7)  &63.53(2)  &3.0(2) &2.08(6) & 4.2\\[-2pt]
G025.04--00.20 MM2$\ast$  &63.63(7)  &63.44(3)  &2.4(2) &2.56(8) & 4.2\\[-2pt]
G025.04--00.20 MM4$\ast$  &63.76(8)  &63.55(3)  &2.0(2) &2.03(6) & 4.1\\[-2pt]
G034.43+00.24 MM1   &58.04(3)  &57.79(4)  &3.85(6)&3.39(9) & 3.5\\[-2pt]
G034.43+00.24 MM2   &57.86(3)  &57.50(4)  &4.22(6)&3.8(1)  & 3.5\\[-2pt]
G034.43+00.24 MM3   &60.05(7)  &58.98(3)  &5.8(2) &2.56(7) & 3.6\\[-2pt]
G034.43+00.24 MM4$\ast$   &57.50(3)  &57.50(4)  &3.22(8)&3.28(9) & 3.5\\[-2pt]
G034.43+00.24 MM5$\ast$   &57.81(8)  &57.57(4)  &1.8(2) &2.3(1)  & 3.5\\[-2pt]
G034.43+00.24 MM6$\ast$   &58.34(6)  &58.31(4)  &2.5(1) &2.59(9) & 3.6\\[-2pt]
G034.43+00.24 MM8   &57.7(2)   &57.84(4)  &4.4(4) &3.14(9) & 3.5\\[-2pt]
G034.43+00.24 MM9$\ast$   &58.40(5)  &58.53(4)  &2.4(1) &2.62(9) & 3.6\\[-2pt]
I18102--1800 MM1    &21.53(5)  &21.53(4)  &4.4(1) &2.88(8) & 2.7\\[-2pt]
I18151--1208 MM1$\ast$    &33.26(5)  &32.84(3)  &2.1(1) &2.37(8) & 2.8\\[-2pt]
I18151--1208 MM2$\ast$    &30.30(1)  &29.93(4)  &3.2(3) &2.9(1)  & 2.6\\[-2pt]
I18182--1433 MM1$\ast$    &59.86(3)  &59.72(4)  &2.58(7)&3.18(9) & 4.6\\[-2pt]
I18223--1243 MM1$\ast$    &45.38(6)  &45.28(3)  &2.1(1) &1.87(7) & 3.6\\[-2pt]
I18223--1243 MM2$\ast$    &45.5(1)   &45.84(5)  &2.1(3) &2.9(1)  & 3.6\\[-2pt]
I18223--1243 MM3$\ast$    &45.65(7)  &45.66(4)  &3.2(2) &2.95(9) & 3.7\\[-2pt]
I18306--0835 MM1    &78.5(1)   &77.88(3)  &4.0(2) &2.57(8) & 4.9\\[-2pt]
I18337--0743 MM1$\ast$    &58.31(4)  &58.37(5)  &3.51(8)&3.7(1)  & 3.9\\[-2pt]
I18337--0743 MM2$\ast$    &58.13(6)  &58.53(7)  &2.3(1) &3.2(2)  & 3.9\\[-2pt]
I18337--0743 MM3$\ast$    &55.6(2)   &55.85(5)  &3.7(4) &3.2(1)  & 3.8\\
\end{tabular}
\end{ruledtabular}
\end{table*}

\subsection{Dense cores in the Pipe Nebula}
\label{Sect3-2}

The second sample of molecular lines is based on high quality data from
Ref.~\cite{RLM08} who observed a starless population of 46
molecular cores in the Pipe Nebula ($D \sim 130$ pc). 
The Pipe Nebula is one of the closest molecular clouds with a mass of 
$10^4 {\cal M}_\odot$ covering an area of 48 deg$^2$ on the sky. 

The observations were performed with the 100-m Robert C. Byrd
Green Bank Telescope (GBT). 
The NH$_3$ (1,1), (2,2), CCS ($2_1-1_0$), and HC$_5$N
(9-8) transitions were observed simultaneously as in the case of the 
Perseus cloud
(the frequency of the HC$_5$N $J=9-8$ transition is 23.96 GHz, and
its upper state excitation temperature 
$E^{\scriptscriptstyle (9,8)}_u = 5.8$ K). 
The GBT beam size of $\sim30''$ provides the angular resolution of 
$\sim0.02$ pc at the Pipe Nebula distance, i.e. the internal structure
of dense cores ($R \simeq 0.09$ pc) was partly resolved.
The velocity resolution was 23 \ms. 

There are four sources in the Pipe Nebula sample (Nos. 12, 40, 101, 109 in
Table~4 in  Ref.~\onlinecite{RLM08}) 
where both NH$_3$ (1,1) and HC$_5$N (9,8) lines
are observed. The molecule HC$_5$N is 4.4 times heavier than NH$_3$, i.e.
its lines should be intrinsically narrower than ammonia lines 
if molecules trace the same gas.  However, each of these sources exhibits 
HC$_5$N velocity dispersion broader than NH$_3$:
in average 
$\sigma_v({\rm HC}_5{\rm N}) = (1.5\pm0.1)\ \sigma_v({\rm NH}_3)$. 
An even worse relationship is seen between HC$_5$N and CCS velocity dispersions:
in average 
$\sigma_v({\rm HC}_5{\rm N}) = (1.8\pm0.5)\ \sigma_v({\rm CCS})$. 
This could be expected if NH$_3$ and CCS  
trace not all the gas along the line of sight
but only a part associated with the dense compact core. 
However, not all of the 12 NH$_3$ and CCS emission sources listed in 
Table~\ref{tbl-3} show the expected correlation of the velocity dispersion.
Four sources marked by open circles in Fig.~\ref{fg3}, upper panel, are
those where two species trace different regions.
The distribution of the remaining 8 points is shown in 
Fig.~\ref{fg3}, lower panel. Again, as for the sources from the Perseus
molecular cloud, we find a positive offset in the radial velocities. 
The difference between the CCS and NH$_3$ velocity
has the mean of ${\Delta V}_{n=8} = 0.044$ \kms\       
and the standard deviation of $0.020$ \kms. 
Adding the error of the rest frequencies, we obtain the final values
${\Delta V}_{n=8} = 0.044 \pm 0.020\pm0.0135$ \kms, and the robust
$M$-estimate
${\Delta V}_{n=8} = 0.069 \pm 0.011\pm0.0135$ \kms\
(Table~\ref{tbl-6}).

%-----------------Figure 8
\begin{figure}[t]
\includegraphics[viewport=19 167 570 720,width=80mm,height=60mm]{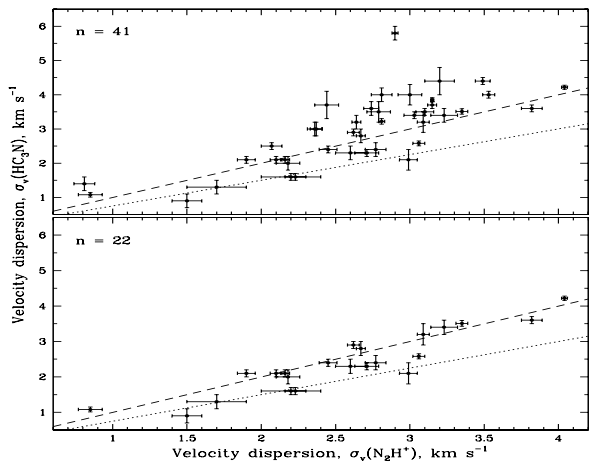}
\caption{\label{fg8}
{\it Upper panel}: 
HC$_3$N ($5-4$) versus N$_2$H$^+$ ($1-0$) linewidths
for massive clumps associated with infrared dark clouds \cite{SSK08}.
The error bars represent $1\sigma$ standard deviations. 
{\it Lower panel}: 
Same as the upper panel but for the points distributed between
the dashed and dotted lines which constrain 
respectively pure turbulent, 
$\sigma_v({{\rm HC}_3{\rm N}}) = \sigma_v({\rm N}_2{\rm H}^+)$,
and pure thermal, 
$\sigma_v({{\rm HC}_3{\rm N}}) = 0.58 \sigma_v({\rm N}_2{\rm H}^+)$,
line broadening. The sample sizes are indicated in the panels.
}
\end{figure}

\subsection{Massive clumps associated with Infrared Dark Clouds}
\label{Sect3-3}

The third example deals with more distant objects~--- cold, dense,
and massive clumps (${\cal M} > 100 {\cal M}_\odot$) of 1-10 pc sizes
associated with the IRDCs which are located at distances larger than 2 kpc and
seen in silhouette against the bright diffuse mid-infrared 
emission of the Galactic plane. 
The gas kinetic temperature of IRDCs is as low as the temperature of 
the dense molecular cores, $T_{\rm kin} < 20$ K \cite{PWC06}.
This means that massive clumps in IRDCs are either starless or in a very
early stage of high-mass star formation. 

%-----------------Figure 9
\begin{figure}[t]
\includegraphics[viewport=19 167 570 720,width=80mm,height=60mm]{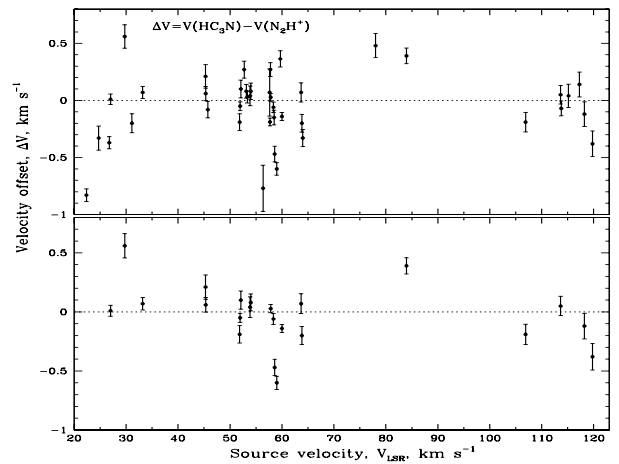}
\caption{\label{fg9}
{\it Upper panel}: 
Velocity offset $\Delta V_{{{\rm HC}_3{\rm N}}-{\rm N}_2{\rm H}^+}$ 
versus the source radial 
velocity set by $V_{\rm LSR}$(N$_2$H$^+$). 
{\it Lower panel}: Same as the upper panel but for the
points from the lower panel in Fig.~\ref{fg8}.
The vertical error bars include both random and systematic errors caused by
the uncertainties of the adopted rest frequencies (Table~\ref{tbl-1}).
}
\end{figure}

A recent survey of 55 massive clumps in NH$_3$ (1,1), (2,2), (3,3),
N$_2$H$^+$ (1-0), and HC$_3$N (5-4) lines has been carried out at the Nobeyama
Radio Observatory 45-m telescope \cite{SSK08}.
The N$_2$H$^+$ and HC$_3$N (5-4) lines (Table~\ref{tbl-1})
were observed simultaneously using two SIS receivers. 
The third receiver HEMT was used for NH$_3$ lines.
The data were obtained within one day for each position,
and different receivers used in these observations 
allow the Doppler tracking to be completely independent between these 
molecular lines \cite{Sa08}.
%(T. Sakai, private communication). 
 
%-------------------------- Table 6
\begin{table*}[t!]
\caption{Sample mean values (unweighted) $\Delta V$, standard deviations
$\sigma_{\rm rms}$, robust $M$-estimates of the sample mean
$\Delta V_{\scriptscriptstyle M}$, and the scale $\sigma_{\scriptscriptstyle M}$  
deduced from the original data and from the subsamples
of molecular lines showing self-consistent linewidths.
Col.~9 lists velocity offsets corrected for uncertainties of the rest frequencies
(see Sect.~\ref{Sect4-1} for more details).
The errors of the mean $\sigma_{\rm rms}/\sqrt{n}$\ and 
$\sigma_{\scriptscriptstyle M}/\sqrt{n}$\ are listed, respectively,
in columns (4) and (6).
}
\begin{ruledtabular}
\label{tbl-6}
\begin{tabular}{cccrcrccr}
\noalign{\smallskip}
 & Molecular & Sample & \multicolumn{1}{c}{$\Delta V$,} 
& $\sigma_{\rm rms}$, & 
\multicolumn{1}{c}{$\Delta V_{\scriptscriptstyle M}$,} &  
$\sigma_{\scriptscriptstyle M}$, & 
$\varepsilon_{{\rm sys},{\Delta V}}$ & 
\multicolumn{1}{c}{$\Delta V^{\rm corr}_{\scriptscriptstyle M}$,} \\
Object & pair & size, $n$ & \multicolumn{1}{c}{\kms} & \kms & 
\multicolumn{1}{c}{\kms} &\kms & \kms & \multicolumn{1}{c}{\kms} \\
${\scriptstyle (1)}$ & ${\scriptstyle (2)}$ & ${\scriptstyle (3)}$ & 
\multicolumn{1}{c}{$\scriptstyle (4)$} & ${\scriptstyle (5)}$ & 
\multicolumn{1}{c}{$\scriptstyle (6)$} & ${\scriptstyle (7)}$ 
& ${\scriptstyle (8)}$ & \multicolumn{1}{c}{${\scriptstyle (9)}$}  \\
\noalign{\smallskip}
\hline
\noalign{\medskip}
Perseus & NH$_3$/CCS & 98 & $0.044\pm0.013$&0.129& $0.040\pm0.007$&0.073& 
0.0135 & 0.024 \\[-2pt] 
        & & 34 & $0.039\pm0.010$&0.058& $0.045\pm0.007$&0.043& 0.0135 & 0.029 \\[-2pt]
        & & 21 & $0.048\pm0.013$&0.060&  $0.052\pm0.007$&0.032& 0.0135 & 0.036 \\[1pt]
Pipe&NH$_3$/CCS&12&$0.087\pm0.039$&0.135&$0.039\pm0.023$&0.081& 0.0135 & 0.023 \\[-2pt]
        & &8&$0.044\pm0.020$&0.057& $0.069\pm0.011$&0.030& 0.0135 & 0.053 \\[1pt]
IRDCs& NH$_3$/N$_2$H$^+$&54&$0.157\pm0.040$&0.294&  
$0.160\pm0.030$&0.220 &0.0136 & 0.148 \\[-2pt]
  & & 36 & $0.122\pm0.049$&0.294&  $0.160\pm0.032$&0.190& 0.0136 & 0.148 \\[1pt]
 &NH$_3$/HC$_3$N&43&$0.138\pm0.043$&0.282& $0.110\pm0.032$&0.210& 0.0031 & 0.105 \\[-2pt]
        & & 27 &$0.105\pm0.045$&0.234& $0.120\pm0.037$&0.190& 0.0031 & 0.115 \\[1pt]
   &N$_2$H$^+$/HC$_3$N&41&$-0.056\pm0.047$&0.301& $-0.020\pm0.037$&0.240& 
0.0138 & $-0.013$ \\[-2pt]
         & & 22 & $-0.033\pm0.055$&0.258& $-0.017\pm0.034$&0.160& 0.0138 & $-0.010$ \\  
\end{tabular}
\end{ruledtabular}
\end{table*}

The angular and velocity resolutions in these observations were different.
At the frequencies of the NH$_3$, HC$_3$N, and N$_2$H$^+$ transitions the
half-power beam width was $73''$,  $37''$, and  $18''$, and
the velocity resolution was 0.50, 0.25, and 0.12 \kms, respectively.
At a distance of $\sim$5 kpc  
the spatial structures of $\sim$1.8, 0.9, and 0.4 pc can
be resolved in, correspondingly, NH$_3$, HC$_3$N, and N$_2$H$^+$ lines.
The inhomogeneity of the observational parameters 
and significant turbulence in IRDC cores 
($\sigma_{\rm turb} \sim 1-3$ \kms) increase the dispersion of the
Doppler noise in massive clumps as compared with low-mass molecular cores.

Table~\ref{tbl-4} lists 54 sources from \cite{SSK08} where both
N$_2$H$^+$ (1-0) and NH$_3$ (1,1) lines were detected. These molecules
are expected to have similar spatial distributions. However,
Fig.~\ref{fg4}, upper panel, 
shows that some of the N$_2$H$^+$ lines have either too large
or too small velocity dispersion as compared to NH$_3$,
which may be related to a large angular resolution difference
($18''$ vs. $73''$) in observations of these molecules.  
Such points were excluded from the sample, the remaining $n=36$ points are
presented in Fig.~\ref{fg4}, lower panel. 
The velocity offset distributions from the total sample and the subsample 
are shown in Fig.~\ref{fg5}, upper and lower panels, respectively.
The weighted and unweighted means are as follows:
${\Delta V}^w_{n=36} = 0.124\pm0.043$ \kms, and
${\Delta V}_{n=36} = 0.122\pm0.049$ \kms.
In this case the laboratory error is equal to 
$\varepsilon_v = 13.6$ \ms. For the
final estimate we take the more appropriate
unweighted mean 
${\Delta V}_{n=36} = 0.122\pm0.049\pm0.0136$ \kms, 
and the robust $M$-estimate
${\Delta V}_{n=36} = 0.160\pm0.032\pm0.0136$ \kms\
(Table~\ref{tbl-6}).

The correlation between velocity dispersions 
of the HC$_3$N (5-4) and NH$_3$ (1,1) emission lines
is shown in Fig.~\ref{fg6}. In this case the results were obtained with
more comparable angular resolutions ($37''$ vs. $73''$).
The total sample of the HC$_3$N/NH$_3$ pairs ($n = 43$) is given
in Table~\ref{tbl-5}. 
Several of the measured HC$_3$N 
linewidths demonstrate significant supersonic
turbulence of the emitting gas with
$\sigma_v$(HC$_3$N) $> \sigma_v$(NH$_3$). 
Such lines were omitted from the following analysis, the
remaining $n = 27$ points are 
shown in Fig.~\ref{fg6}, lower panel (the selected points are
marked by asterisks in Table~\ref{tbl-5}, first column). 
Fig.~\ref{fg7} illustrates the distribution of the measured offsets. 
The calculated statistics are as
follows: 
${\Delta V}^w_{n=27} = 0.147\pm0.039$ \kms, and
${\Delta V}_{n=27} = 0.105\pm0.045$ \kms.
Again, taking the unweighted estimate 
and accounting for the systematic error of
$\varepsilon_v = 3.1$ \ms\ 
we come to the final mean value of the velocity offset
${\Delta V}_{n=27} = 0.105\pm0.045\pm0.0031$ \kms, and
the robust $M$-estimate
${\Delta V}_{n=27} = 0.120\pm0.037\pm0.0031$ \kms\
(Table~\ref{tbl-6}). 

%-------------------------- Table 7
\begin{table*}[t!]
\caption{Hyperfine components of the HC$_3$N $J = 5-4$ transition.
For M\"oller (Ref.~\onlinecite{MSS05}), Yamamoto (Ref.~\onlinecite{YSK90}), and
JPL frequencies only the numbers after the decimal point are shown, 
the integer part is as
in Lapinov's data \cite{Lap08}. Col.~10 and 11 list velocity 
offsets between Yamamoto and Lapinov (Y--L) 
and between JPL and Lapinov  (JPL--L) values.
}
\begin{ruledtabular}
\label{tbl-7}
\begin{tabular}{cccclclcccc}
\noalign{\smallskip}
\multicolumn{4}{c}{Transition} &  &
\multicolumn{4}{c}{ Rest Frequency (MHz)} &
\multicolumn{2}{c}{ $\Delta v$ (\ms) }\\
$J'$ & $F'$ & $J$ & $F$ & \multicolumn{1}{c}{Weight} & Lapinov & 
\multicolumn{1}{c}{M\"oller} & Yamamoto & JPL & Y--L & JPL--L \\
${\scriptstyle (1)}$ & ${\scriptstyle (2)}$ & ${\scriptstyle (3)}$ & 
\multicolumn{1}{c}{$\scriptstyle (4)$} & 
\multicolumn{1}{c}{${\scriptstyle (5)}$} & 
\multicolumn{1}{c}{$\scriptstyle (6)$} & 
\multicolumn{1}{c}{${\scriptstyle (7)}$} 
& ${\scriptstyle (8)}$ & ${\scriptstyle (9)}$ & ${\scriptstyle (10)}$ &
${\scriptstyle (11)}$
\\
\noalign{\smallskip}
\hline
\noalign{\medskip}
5&5&4&5& 0.034 & 45488.83643(42) & .8368(10) &  & & & \\[-2pt]
5&4&4&3& 0.658 & 45490.26138(42) & .2614(5) & .263(1) & .2580(3) & 10.7 & $-22.3$ \\[-2pt]
5&5&4&4& 0.812 & 45490.31378(42) & .3137(5) & .316(1) & .3102(3) & 14.6 & $-23.6$ \\[-2pt]
5&6&4&5& 1.000 & 45490.33740(42) & .3373(6) & .341(1) & .3336(4) & 23.7 & $-25.0$ \\[-2pt]
5&4&4&5& 0.0003 & 45490.63123(43) & .6317(16) &  &  & & \\[-2pt]
5&4&4&4& 0.034 & 45492.10858(42) & .1085(9) &  &  & & \\
\end{tabular}
\end{ruledtabular}
\end{table*}

The data from Ref.~\cite{SSK08} allow us to answer 
the question whether the revealed
positive velocity offsets are related to ammonia lines or not. Following
the same procedure as above, we can compare the velocities obtained from the
sources which were observed in both N$_2$H$^+$ and HC$_3$N.
The total sample of $n = 41$ pairs is shown in Fig.~\ref{fg8}, 
upper panel.  The heavier molecule HC$_3$N should have 
linewidths smaller or equal to that of N$_2$H$^+$.
However, a considerable number of points ($\sim$50\%) lie above the dashed line
constraining pure turbulent line broadening. This is expected since 
N$_2$H$^+$ molecule traces more compact central regions of the dense clouds. 
The $n = 22$ points which
presumably trace the same gas are shown in Fig.~\ref{fg8}, lower panel. 
The corresponding velocity offset distributions are shown in Fig.~\ref{fg9}.
The mean values are  
${\Delta V}^w_{n=22} = -0.053\pm0.044$ \kms, and
${\Delta V}_{n=22} = -0.033\pm0.055$ \kms, and thus
the final estimates are
${\Delta V}_{n=22} = -0.033\pm0.055\pm0.0138$ \kms, and
the robust $M$-value
${\Delta V}_{n=22} = -0.017\pm0.034\pm0.0138$ \kms\
(Table~\ref{tbl-6}).

In this case 
both weighted and unweighted values are consistent with zero velocity
offset. Since N$_2$H$^+$ and NH$_3$ share similar physical
conditions in their formation and destruction processes and, hence,  trace
each other in spatial distributions, the obtained result indicates
that ammonia is a key element responsible for the revealed velocity offset. 

To demonstrate that the positive velocity offset does not result from 
our selection criterion we summarize all measurements
complemented by robust redescending $M$-estimates for the shift and scale
in Table~\ref{tbl-6}. 
Both results are in good agreement within the $1\sigma$
uncertainty interval.
The sample mean values reproduce a marginally
significant signal ($\ga 3\sigma$)
independently on the sample size. 
When ammonia lines are excluded from the analysis, the signal disappears.

\section{Discussion}
\label{Sect4}

In the previous section we showed that
different molecular clouds demonstrate statistically significant
velocity offset between the ammonia inversion transition $(J,K) = (1,1)$
and rotational transitions of other molecules.  
Here we consider possible reasons of this offset.

\subsection{Uncertainties of the rest frequencies}
\label{Sect4-1}

In cold dark clouds with the kinetic temperature $T_{\rm kin} \sim 10$K the
thermal line broadening is very small ($v_{\rm th} \sim 50$ \ms\ for, e.g., CCS).
This means that the detailed study of velocity offsets requires accuracy of
rest frequencies of about 1 \ms\ \cite{La06}.
Among molecules used in the
present work only NH$_3$ fulfills this requirement. 
The hyperfine structure of the observed
$(J,K) = (1,1)$ transition of NH$_3$
is known with the accuracy of $\simeq 0.6$ \ms\ \cite{Ku67}.
The available laboratory uncertainties for 
other molecules are significantly larger.
Besides, there are shifts between rest frequencies 
reported in different publications.
Below we estimate such systematics and their effect on 
the mean values of the velocity offsets obtained in Sect.~\ref{Sect3}.

The frequency 22.344033 GHz of the CCS $J_N = 2_1 - 1_0$ 
transition (Table~\ref{tbl-1})
used in Refs.~\cite{RPF08,RLM08} 
was calculated from a comparison with the three strongest 
hyperfine components of the HC$_3$N $J = 5-4$ transition
observed towards a cold dark cloud L1498 assuming that the spatial distribution
of CCS is the same as that of HC$_3$N \cite{YSK90}.
According to Ref.~\cite{YSK90}, the $V_{\rm LSR}$ value obtained from the HC$_3$N
transition was used to determine the frequencies of the CCS transitions without
any corrections, and the frequencies of the CCS transitions were calculated from
Gaussian fitting of the observed profiles. 
The error of 1 kHz indicated in Table~\ref{tbl-1}
corresponds to one standard deviation obtained in the fit.
However, up to now the $J = 5-4$ transition has not been measured in laboratory. 
The analysis of all available laboratory data on HC$_3$N carried out 
in Refs.~\cite{MTR01,MSS05}
and independently in Ref.~\cite{Lap08}
%by A. V. Lapinov (private communication) 
shows systematic frequency shifts (Table~\ref{tbl-7}) with respect to the
data presented in Ref.~\cite{YSK90} and in the JPL 
Catalog (http://spec.jpl.nasa/gov/). 
On the other hand both Ref.~\cite{Lap08} and 
Ref.~\cite{MSS05} frequencies are in good agreement 
with the former having slightly better accuracy ($\varepsilon_v \simeq 2.8$ \ms). 
The mean value for the shift between 
rest frequencies used in Ref.~\cite{YSK90} and
calculated in Ref.~\cite{Lap08}  is $\Delta V_{\rm shift} = 16$ \ms\
(Col.~10, Table~\ref{tbl-7}). Thus, the presented in Table~\ref{tbl-6} 
robust $M$-estimates of the velocity offsets
between CCS and NH$_3$ towards
the Perseus molecular cloud and the Pipe Nebula corrected 
for this systematic shift will be
$\Delta V^{\rm corr}_{n=21} = \Delta V_{n=21} - \Delta V_{\rm shift} = 
0.036\pm0.007\pm0.0135$ \kms, and
$\Delta V^{\rm corr}_{n=8} = \Delta V_{n=8} - \Delta V_{\rm shift} = 
0.053\pm0.011\pm0.0135$ \kms, respectively (Table~\ref{tbl-6}, Col.~9).

%-------------------------- Table 8
\begin{table*}[t!]
\caption{Hyperfine components of the N$_2$H$^+$ $J = 1-0$ transition.   
For Caselli (Ref.~\onlinecite{CMT95})
frequencies only the numbers after the decimal point are 
shown, the integer part is as
in the M\"oller data (Ref.~\onlinecite{MSS05}). Col.~10 lists velocity
offsets between Caselli and M\"oller (C--M) values.
}
\begin{ruledtabular}
\label{tbl-8}
\begin{tabular}{cccccccccc}
\noalign{\smallskip}
\multicolumn{6}{c}{Transition} &  &
\multicolumn{2}{c}{ Rest Frequency (MHz)$^\dagger$ } &
\multicolumn{1}{c}{ $\Delta v$ (\ms) }\\
$J'$ & $F'_1$ & $F'$ & $J$ & $F_1$ & $F$ & \multicolumn{1}{c}{Weight} & 
\multicolumn{1}{c}{M\"oller} & Caselli & C--M \\
${\scriptstyle (1)}$ & ${\scriptstyle (2)}$ & ${\scriptstyle (3)}$ & 
\multicolumn{1}{c}{$\scriptstyle (4)$} & 
\multicolumn{1}{c}{${\scriptstyle (5)}$} & 
\multicolumn{1}{c}{$\scriptstyle (6)$} & 
\multicolumn{1}{c}{${\scriptstyle (7)}$} 
& ${\scriptstyle (8)}$ & ${\scriptstyle (9)}$ & 
\multicolumn{1}{c}{${\scriptstyle (10)}$}  \\
\noalign{\smallskip}
\hline
\noalign{\medskip}
1&1&0 &0&1&1& 0.143 & 93171.6171(41) & .6210(70) &  12.5 \\[-2pt]
1&1&2 &0&1&2& 0.604 & 93171.9129(42) &  .9168(70) & 12.5  \\[-2pt]
1&1&2 &0&1&1& 0.111 & 93171.9129(42) &  &   \\[-2pt]
1&1&1 &0&1&0& 0.218 & 93172.0499(42) & .0533(70) & 10.9  \\[-2pt]
1&1&1 &0&1&2& 0.159 & 93172.0499(42) &  &   \\[-2pt]
1&1&1 &0&1&1& 0.051 & 93172.0499(42) &  &   \\[-2pt]
1&2&2 &0&1&1& 0.604 & 93173.4748(42) & .4796(70) & 15.4  \\[-2pt]
1&2&2 &0&1&2& 0.111 & 93173.4748(42) &  &   \\[-2pt]
1&2&3 &0&1&2& 1.000 & 93173.7722(41) & .7767(70) & 14.5  \\[-2pt]
1&2&1 &0&1&1& 0.278 & 93173.9657(43) & .9666(70) &  2.9 \\[-2pt]
1&2&1 &0&1&0& 0.134 & 93173.9657(43) &  &   \\[-2pt]
1&2&1 &0&1&2& 0.017 & 93173.9657(43) &  &   \\[-2pt]
1&0&1 &0&1&2& 0.252 & 93176.2611(43) & .2650(70) & 12.5   \\[-2pt]
1&0&1 &0&1&1& 0.100 & 93176.2611(43) &  &   \\[-2pt]
1&0&1 &0&1&0& 0.076 & 93176.2611(43) &  &   \\
\end{tabular}
\end{ruledtabular}
\end{table*}

Now let us consider the IRDCs from Ref.~\cite{SSK08}. 
In this measurements the lower spectral resolution does not
allow to resolve hyperfine structure of the HC$_3$N $J=5-4$ transition.
The spectra of the HC$_3$N lines were fitted to a single
Gaussian function and the corresponding radial velocities
were calculated on base of a single  
hyperfine line $J=5-4, F=5-4$\ \cite{Sa08}\ 
%(T. Sakai, private communication) 
with the frequency 45490.3102 MHz taken from the JPL Catalog
(Table~\ref{tbl-7}, Col.~9).
As a result, a small systematic shift due to unresolved hyperfine components
was introduced. Its value can be estimated if we consider
the center of gravity for the weighted central hyperfine 
components of the $J=5-4$ transition
listed in Table~\ref{tbl-7}. The frequency of the center of gravity is
$\nu_{\rm c.g.} = 45490.3058$ MHz and thus
$\Delta V_{\rm c.g.} = 29$ \ms.
However, as seen from Table~\ref{tbl-7} (Col.~9), 
the JPL frequencies are systematically shifted 
with respect to Ref.~\cite{Lap08} data and the mean value of the shift is
$\Delta V_{\rm shift} = -23.6$ \ms\ (Col.~11). 
Combining both shifts, we obtain a correction of 5.4 \ms.
This means that the velocity offset between 
HC$_3$N and NH$_3$ listed in Table~\ref{tbl-6}
should be as follows:
$\Delta V^{\rm corr}_{n=27} = 0.115\pm0.037\pm0.0031$ \kms\
(Table~\ref{tbl-6}, Col.~9).  
    
The rest frequencies for the second molecule N$_2$H$^+$ 
observed towards the IRDCs 
Ref.~\cite{SSK08} were taken from Ref.~\cite{CMT95}. The hyperfine
frequencies of the N$_2$H$^+$ $J=1-0$ transition were estimated from observations
of a molecular cloud L1512 in Taurus in both  N$_2$H$^+$ (93 GHz)
and C$_3$H$_2$ (85 GHz) lines assuming that the C$_3$H$_2$ and the N$_2$H$^+$
have exactly the same radial velocity. 
The absolute frequencies of the  N$_2$H$^+$ hyperfine 
transitions measured in Ref.~\cite{CMT95}
are known with precision of $\sim 7$ kHz. 
However, a factor of two more accurate data are given in 
the Cologne Database for Molecular Spectroscopy (CDMS) described in
Refs.~\cite{MTR01,MSS05}. Both datasets are  compared in Table~\ref{tbl-8}.
The error of 4 kHz ($\varepsilon_v \simeq 13.5$ \ms)
reported in Table~\ref{tbl-1} corresponds to the CDMS. 
The mean shift between the data from Ref.~\cite{CMT95} and the CDMS frequencies is
$\Delta V_{\rm shift} = 12$ \ms. Thus, the velocity offset between
N$_2$H$^+$ and NH$_3$ from Table~\ref{tbl-6} should be corrected to
$\Delta V^{\rm corr}_{n=36} = \Delta V_{n=36} - 0.012 =
0.148\pm0.032\pm0.0136$ \kms,  
and between HC$_3$N and N$_2$H$^+$ to
$\Delta V^{\rm corr}_{n=22} = \Delta V_{n=22} + 0.0066 =
-0.010\pm0.034\pm0.0138$ \kms\ (Table~\ref{tbl-6}, Col.~9).

\subsection{Systematics due to kinematic shifts}
\label{Sect4-2}

Spectroscopic measurement of the absolute radial velocities at the level of 
$50-100$ \ms\ is subject to many physical effects \cite{LD03}.
Fortunately, relative radial velocities are less sensitive to most of them.
However, some small systematic shifts, 
in addition to those discussed in Sect.~\ref{Sect2}, 
may be present in the measured values of $\Delta V$.

First of all, the measurement of the relative radial
velocities between two molecules requires the use of a common reference frame. 
In radio-astronomical observations 
it is usually the Local Standard of Rest (LSR). 
The transformation of the apparent Doppler velocity to $V_{\rm LSR}$
introduces an error of $\sim 1-3$ \ms, stemming from the following corrections: 
the observer's motion due to ($1$) the Sun's motion with respect to the
LSR, ($2$) revolution of the Earth-Moon barycenter round the Sun,
($3$) motion of the Earth center round the  Earth-Moon barycenter, and
($4$) rotation of the Earth \cite{Go76}. This error ought to 
be added to the total error budget along with the 
systematic errors listed in Table~\ref{tbl-6}, 8$th$ column
if molecular lines were not observed simultaneously.

The second error arises from gravitational shifts of the spectral lines
due to the annual variation of the observer's distance from the Sun because of
the eccentricity ($e \simeq 0.01671$)
of the Earth's orbit ($A = 1$ AU). This 
produces the peak-to-peak variation of the radial velocity
(in units of the speed of light, $c$) of
\begin{equation}
\delta V_{\rm g} = \frac{2e G {\cal M}_\odot}{A c^2}\, . 
\label{S4eq2}
\end{equation}
The numerical value of $\delta V_{\rm g}$ is about 10 \cms, 
and in our case this error can be neglected.

If we further 
define the dimensionless gravitational potential of the Sun on
the Earth's orbit as (negative sign is omitted)
\begin{equation} 
\Phi_0 = G {\cal M}_\odot/(A c^2) + G {\cal M}_\oplus/(R_\oplus c^2)
\simeq 1.06\times10^{-8}\, , 
\label{S4eq3}
\end{equation}
then the
gravitational redshift for photons escaping an object of mass ${\cal M}$ 
and radius $R$ scales as
\begin{equation}
\delta V_{\rm g} = 
\left(\frac{\cal M}{{\cal M}_\odot}\right)
\left(\frac{A}{R}\right)\Phi_0\ . 
\label{S4eq4}
\end{equation}

Different gravitational redshifts between molecular lines
may occur from the chemical stratification in dense molecular cores.
However, this effect is negligible because of a relatively
low gas density and rather large linear sizes of the clouds.
For a typical starless core shown in Fig.~3 in Ref.~\cite{DFE07},
the gravitational velocity shifts of 
NH$_3$ and CCS lines are $\sim0.2$ \cms\ and $\sim 0.1$ \cms, respectively.

Systematically redshifted CCS $2_1-1_0$ 
lines with respect to the NH$_3$ (1,1) central velocity may occur
in contracting clouds when the near side of
the external shell of the cloud with the CCS emission is moving towards
the higher density core as suggested for, e.g, the L1551 dark cloud \cite{SWD05}.
However, as discussed in Sect.~\ref{Sect2}, in dark clouds
both kinds of kinematic shifts (redward and blueward) are observed. 
Such bulk motions should randomize the Doppler shifts and provide a zero offset
for a large size sample of dark clouds.
Besides, all molecular cores in the Perseus molecular cloud are probably
optically thin in the CCS $2_1-1_0$ emission since 
CC$^{34}$S was not detected along any of the lines-of-sight \cite{RPF08} and
$^{32}$S : $^{34}$S = 22 : 1.
This means that the CCS emission line profile is formed in the whole volume of the
cloud including both its near and distant sides. 
This is not the case, however, for the NH$_3$ emission. The total
opacity in the (1,1) line transition varies between 0.84 and 10.9 for
the cores listed in Table~\ref{tbl-2}. The optically thick clouds may
cause some systematics of both signs (positive or negative depending on
the character of the internal bulk motions within the core) 
in the central positions of the NH$_3$ lines. However, three cores with
$\tau_{\scriptscriptstyle (1,1)} \la 1$ 
(Nos. 20, 112, and 151) demonstrate 
positive velocity offsets with the mean value of $\sim$100 \ms.  

Among analyzed clouds the most accurate measurements of $\Delta V$
belong to the Perseus cores. 
We may use the $M$-estimate from the $n = 21$ sample
(Sect.~\ref{Sect4-1})
as a tentative value for the found velocity offset, 
$\Delta V = 36 \pm 7_{\rm stat} \pm 13.5_{\rm sys}$ \ms, 
or $\Delta V = 36 \pm 15$ \ms\ ($1\sigma$ C.L.).
Thus, the considered systematics do not explain the revealed offset.
Assuming that this offset is caused entirely by the change of $\mu$,
one obtains from Eq.(\ref{S2eq3})\ 
$\Delta \mu/\mu = (3.5\pm1.4)\times10^{-8}$,    
or $|\Delta \mu/\mu| < 5\times10^{-8}$.
This is the most accurate estimate of \dmm\ 
based on the spectral analysis of astronomical objects. 
Next we consider whether this value is in line with other constraints.

%-----------------Figure 10
\begin{figure}[t]
\includegraphics[viewport=19 167 570 720,width=60mm,height=60mm]{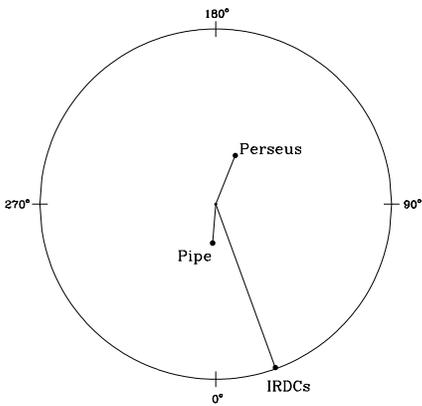}
\caption{\label{fg10}
Schematic location of the Perseus molecular cloud 
($\ell \sim 158^\circ$, $b \sim -20^\circ$, $D \sim 260$ pc), the Pipe Nebula
($\ell \sim 356^\circ$, $b \sim 7^\circ$, $D \sim 130$ pc), and the IRDCs
($\ell \sim 15^\circ-30^\circ$, $b < 1^\circ$, $D > 2$ kpc) in projection
onto the Galactic plane. The Galactic center longitude is 
$\ell = 0^\circ$, by definition. Note that the distance to the
IRDCs is not scaled and they lay much further out.
}
\end{figure}

\subsection{Gravitational-potential dependence of $\mu$}
\label{Sect4-3}

First we explore the dependence of the electron and proton masses
on the strength of the gravitational potential.
Theoretical models dealing with variations of physical
constants do not require the cosmological
temporal evolution of $\alpha$ and/or $\mu$ to be dominated over their local
spatial changes in the Milky Way. 
For instance, theories like that of Ref.~\cite{Be82}
consider direct coupling between the local matter density and
the scalar fields that drive varying constants.
The spatial change of the constants in these models 
is proportional to the local gravitational potential and 
the rate of spatial variation may exceed the rate of time variation
(see, e.g., Ref.~\onlinecite{OP08} and references cited therein).
The effect of scalar fields on $\mu$ is additive (similar to
gravity) and can be expressed as
(Refs.~\onlinecite{F07,FS08}): 
\begin{equation}
\frac{\Delta \mu}{\mu} = k_\mu \Delta \Phi\ ,
\label{S4eq5}
\end{equation}
where $k_\mu$ is a dimensionless coupling constant of a massless (very light)
scalar field to the local gravitational potential $\Phi$, and
$\Delta \Phi$ is the difference of the gravitational potentials
between two measurement points.

This class of models suggests that the coupling of the scalar field 
$\varphi$ to baryonic matter and fields is not 
too strong since otherwise it would produce a 
scalar-mediated `fifth force' that is not observed \cite{GK04}.
It is also assumed that the $\varphi$-mediated force is a long-range one.

The respective contribution from the Galactic gravitational potential 
$\Phi_{\scriptscriptstyle \rm G}(r,\zeta)$ and the
molecular cloud gravitational potential $\Phi_{\rm cloud}$ 
to the local gravitational potential is the sum
\begin{equation}
{\Phi}(r,\zeta) = 
\Phi_{\scriptscriptstyle \rm G}(r,\zeta) + \Phi_{\rm cloud}\ ,
\label{S4eq6}
\end{equation}
where $r$ and $\zeta$ denote two of the cylindrical coordinates.

The gravitational potential of IRDCs 
(${\cal M} \sim 100{\cal M}_\odot$, $R \sim 1$ pc)
and starless cores 
(${\cal M} \sim 10{\cal M}_\odot$,  $R \sim 0.1$ pc) 
is much smaller than that of the Sun on the Earth's orbit,
$\Phi_0$, namely, $\Phi_{\rm cloud} \sim 1/2000\ \Phi_0$.
The value of 
${\Phi}_{\scriptscriptstyle \rm G}(r,\zeta)$ can be calculated from
analytic models of the Milky Way \cite{MN75,MSO80,NMM06} 
which describe fairly well the galactic rotation curves. 
The galactic coordinates of the analyzed clouds are shown 
in Fig.~\ref{fg10}. The distance
of the Sun from the center of the Milky Way 
is $r_0 = 7.94\pm0.42$ kpc\ \cite{ESG03}. 
The Perseus molecular cloud is located towards the galactic anti-center, whereas 
the Pipe Nebula and the IRDCs in the direction of the galactic center.
Between 3 and 8 kpc the galactic potential is about $2\times10^{-6}$, and 
${\Phi}_{\scriptscriptstyle \rm G}$ 
changes within this interval very slowly: 
$\Delta {\Phi}_{\scriptscriptstyle \rm G} \sim 10^{-7}$. 

With \dmm = $3.5\times10^{-8}$,   one obtains $k_\mu \sim 1$, 
which is in sharp contradiction with estimates based on 
atomic clock experiments:
$k_\mu = (-1.1\pm1.7)\times 10^{-5}$ \cite{BLC08}.
There are  two possible explanations for such a large discrepancy: 
\begin{enumerate}
\item The measured
velocity offset is mainly due to kinematic shifts. 
Then, taking the atomic clock limit
$|k_\mu| < 3\times 10^{-5}$
and the change of the galactic gravitational
potential $\Delta \Phi \sim 10^{-7}$, one obtains from
Eq.~(\ref{S4eq5}) an upper limit on 
$|\Delta \mu/\mu| \la 10^{-12}$.
Unfortunately, 
the level of $10^{-12}$ is below the precision of modern astronomical
observations of molecular lines.
\item The electron-to-proton mass ratio does not 
linearly correlate with metric variations,
i.e., $k_\mu \equiv 0$ in Eq.~(\ref{S4eq5}). The nonzero velocity
offset is then caused by the change of the electron and proton masses.
The corresponding theoretical background for this item
is discussed below.
\end{enumerate}

\subsection{Dependence of $\mu$ on matter density}
\label{Sect4-4}

A class of models where baryon masses and coupling constants are
strongly dependent on the local matter density, $\rho$,
and where spatial variations of $\mu(\rho)$ and $\alpha(\rho)$
can be more pronounced than the temporal cosmological variations
was considered in a number of publications discussed in Ref.~\cite{OP08}. 
In this approach it is suggested that the coupling of a scalar field to matter is
much stronger than gravitational, but the fifth force problem does not 
arise since the linear coupling of the scalar field to matter 
is effectively suppressed by the matter density itself.
The dynamics of the scalar field in these models depends on $\rho$:
in low-density environments it is determined by the scalar field potential
$V(\varphi)$, whereas in high-density regions (like the Earth's surface)
it is set by the matter-$\varphi$ coupling. The range of the scalar-mediated
force for the terrestrial matter densities is then very short, less than 1 mm
\cite{OP08}. 

Because of the strong $\varphi$-coupling to matter 
the measurements of the frequency drifts in the
atomic clock experiments ought to be insensitive to the changes in the
gravitational potential at Earth caused by the eccentricity of Earth's orbit.
The minimum of $V(\varphi)$
would show no shifts and \daa\ (\dmm) would be zero with good accuracy.
However, comparison of atomic clocks at high altitude satellite orbits
with laboratory clocks may yield useful constraints on model parameters as
discussed in \cite{OP08} who also suggested to 
test spatial variations of the physical constants 
through atomic and molecular line observations of the interstellar 
clouds in our galaxy. 
In this case the difference of matter densities between 
the terrestrial environment 
($\rho_\oplus \sim 3\times10^{24}$ GeV~\cmm) and, e.g., dense molecular cores 
($\rho_{\rm cloud} \sim 3\times10^{5}$ GeV~\cmm) may cause a
shift in the expectation value of the scalar field
$\varphi_m$ leading to a change in masses and coupling constants.

If we interpret the nonzero offset of $\Delta V \sim 40$ \ms\  in terms of 
the change of fermion masses, then 
for low density environments, such as the interstellar clouds, the change in
the electron-to-proton mass ratio is given by \cite{OP08}:
\begin{equation}
\frac{\Delta \mu}{\mu} \simeq 
(\xi_{\rm e} - \xi_{\rm p})\ \frac{\varphi^2_m}{2}\ .
\label{S4eq7}
\end{equation}
In this model the scalar field is coupled to the standard model fermions 
through the dimensionless constants 
$\xi_{\rm e}$ and $\xi_{\rm p}$, with $\xi_{\rm p}$ being normalized to one. 

The factor $(\xi_{\rm e} - 1)$ can be much larger than unity in accord with
astrophysical constraints discussed in \cite{OP08}.
If we assume that $(\xi_{\rm e} - 1)/2 \ga 1$, then
$\varphi^2_m \la 3.5\times10^{-8}$, or $\varphi_m \la 2\times10^{-4}$.  

Thus, considered observations of the interstellar molecular clouds
being interpreted in terms of the \dmm\ variation
are in agreement with models which treat the $\varphi$-mediated
force as a short-range force depending on the matter density.
In accord with model predictions, the estimated
value of the scalar field is much less than one.  
The measured spatial variation of \dmm\ does
not contradict the laboratory studies on atomic clocks (see Sect.~\ref{Sect4-3})
since extremely different density environment
in the terrestrial measurements and in the interstellar medium
prohibits direct comparison of the obtained results. 

There are several extragalactic measurements of \dmm\ in which gas densities 
similar to those in the Milky Way
clouds are observed. These results, being interpreted in
terms of the spatial variations of \dmm, 
can be compared with our estimate of
$\Delta \mu/\mu = (3.5\pm1.4)\times10^{-8}$: 
it is consistent with the limit on
$|\Delta \mu/\mu| < 3\times10^{-6}$ derived from the NH$_3$-bearing
clouds at $z = 0.68$ \cite{FK07b}, but in conflict with the value
$\Delta \mu/\mu = (-24\pm6)\times10^{-6}$ obtained from molecular
hydrogen H$_2$ absorption lines at $z \sim 3$ \cite{RBH06}.
However, the same H$_2$-bearing clouds at $z \sim 3$ 
as analyzed in Ref.~\cite{RBH06}
were studied in Ref.~\cite{WR08}. 
The variability of $\mu$ was not confirmed at the level
of $|\Delta \mu/\mu| \leq 4.9\times10^{-5}$.
Recently a more stringent limit on \dmm\ was found 
at $z \sim 3$ in Ref.~\cite{KWM08}:
$\Delta \mu/\mu = (2.6\pm3.0)\times10^{-6}$.
Since at densities as low as observed in the interstellar medium
the dependence of $\mu$ on $\rho$ is extremely weak, the value of \dmm\
in quasar absorbers is expected to be at the same level as
in the interstellar clouds, i.e. $\sim 10^{-8}-10^{-7}$. 
This level of accuracy can be probed with the next
generation of high resolution spectrographs such as ESPRESSO planned for
the Very Large Telescope \cite{Mo07}.

\subsection{Temporal vs spatial variability of $\mu$}
\label{Sect4-5}

Our measurements of \dmm\ being formally divided by the distance to the sources
in light years  imply a time variation of 
\dmm\ of about $10^{-11}-10^{-10}$ yr$^{-1}$. 
This variation rate is orders of magnitude 
larger than the laboratory bound quoted in Sect.~\ref{Sect1},
$|\dot{\mu}/\mu| < 10^{-13}$ yr$^{-1}$, and than the
extragalactic bounds based on the upper limit on \dmm\ at $z \sim 3$, 
$|\dot{\mu}/\mu| < 6\times10^{-16}$ yr$^{-1}$\, \cite{KWM08},
and at $z = 0.68$,
$|\dot{\mu}/\mu| < 4\times10^{-16}$ yr$^{-1}$\, \cite{FK07b}.
Such glaring discrepancy may serve as another argument in favor of chameleon-type
models: due to principal dependence of couplings and masses
on matter density the laboratory bounds and
astronomical measurements cannot be directly compared.  
Taken at face values,
we may suggest that the rate of the spatial variations of $\mu$ exceeds 
significantly the rate of temporal variations.

\section{Conclusions}
\label{Sect5}

In this paper we propose 
to use the ammonia inversion transitions in conjunction with
low-lying rotational transitions of other molecules
to probe the spatial changes of the electron-to-proton mass ratio $\mu$
from observations of molecular clouds located in the disk of the Milky Way
at different galactocentric distances.

The reported estimates of \dmm\ are obtained for the Perseus molecular
cores \cite{RPF08}, the Pipe Nebula \cite{RLM08},
and the infrared dark clouds \cite{SSK08}.
We found evidence for the spatial variability of $\mu$ at the level of 
$\Delta\mu/\mu \sim (4-14)\times10^{-8}$ 
which is the most accurate astrophysical value to date. 
The statistical reliability of this result
seems to be high enough but some instrument
systematics could be possible. The revealed positive
velocity offset between the NH$_3$ inversion transition and
rotational transitions of other molecules should
be confirmed by independent observations at different
radio telescopes. 

The result obtained can be compared with
other tests of spatial variations of physical constants
in the solar system which are based on atomic clock laboratory
measurements. Thus, 
if \dmm\ follows the gradient of the local
gravitational potential, as suggested in some scalar field models,
then our estimate of \dmm\ 
contradicts significantly ($\sim 5$ orders of magnitude)
the atomic clock constraints obtained
at different points in the gravitational potential of the Sun
on the Earth's orbit.
However, the measured signal of
\dmm $\sim (4-14)\times10^{-8}$
is in agreement with chameleon-type scalar
field models which predict a strong dependence of masses and coupling
constants on the ambient matter density.

To be completely confident 
that the derived velocity shift is not due to kinematic effects in the
clouds but is the appearance of the spatial variation of \dmm,
new high precision radio-astronomical
observations are needed for a wider range of objects.
Such observations should include
essentially optically thin lines of
molecules co-spatially distributed with ammonia in order to
reduce the Doppler noise. 
It is also desirable to increase  
the accuracy of the rest frequencies
of the CCS ($2_1-1_0$) and N$_2$H$^+$ ($1-0$) transitions 
since their present uncertainties translate
into the radial velocity error of $\sim 13$ \ms\  whereas the error of the
velocity shift between, for instance, CCS and NH$_3$ in the
Perseus cloud can be 
as small as $5$ \ms\ (Table~\ref{tbl-2}). 

In addition, the search for spatial variations of the fine-structure
constant $\alpha$ in the Milky Way using mid- and 
far-infrared fine-structure transitions in atoms and ions \cite{KPL08}, 
or the search for spatial variations of the combination of $\alpha^2/\mu$
using the [{C}\,{\sc ii}] $\lambda158$ $\mu$m line and 
CO rotational lines \cite{LRK08}
would be of great importance for cross-checking the results.

\begin{acknowledgements}
The authors are grateful to
Eric Rosolowsky, Takeshi Sakai, and Jill Rathborne
who sent us additional comments on their observations
and data reduction.
We thank Paola Caselli, Irina Agafonova, Keith Olive, and Thomas Dent 
for useful comments and discussions on the topic, and
Alexander Lapinov for computed frequencies of the HC$_3$N $J=5-4$ 
hyperfine transitions on which Table~\ref{tbl-7} and Sect.~\ref{Sect4-1}
are based.
This work is supported by
the RFBR grants 06-02-16489, 07-02-00210, 08-02-00460,
and by the Program `Leading Scientific Schools of RF'
(grant NSh-2600.2008.2).
\end{acknowledgements}

\end{document}